%% file: main.tex
\documentclass[sigconf, nonacm]{acmart}

\usepackage{graphicx}
\usepackage{xspace}
\usepackage[vlined, ruled, linesnumbered]{algorithm2e}
\usepackage{multirow}
\usepackage[shortlabels]{enumitem}
\usepackage{subcaption}
\usepackage[skins, fitting]{tcolorbox}
\usepackage{balance}
\usepackage{environ}
\usepackage{xcolor}
\usepackage{amsmath} 
\usepackage[normalem]{ulem}

\def\naive{na\"{\i}ve\xspace}

\newcommand{\stitle}[1]{\vspace{0.2em}\noindent\textbf{#1.}}

\newcommand{\systitle}{PathFinder}
\newcommand{\sys}{PathFinder\xspace}

\newcommand{\utility}{search utility\xspace}

\newcommand{\pgraph}{proximity graph\xspace}

\newcommand{\gsp}{graph search plan\xspace}
\newcommand{\btree}{B\textsuperscript{+}-tree\xspace}

\newcommand{\prefilter}{Pre-filtering\xspace}
\newcommand{\infilter}{In-filtering Vamana\xspace}
\newcommand{\oor}{OOR Vamana\xspace}
\newcommand{\acorn}{ACORN\xspace}
\newcommand{\randomselect}{RandomSelect\xspace}

\newcommand{\arxiv}{ArXiv\xspace}
\newcommand{\redcaps}{RedCaps\xspace}
\newcommand{\sift}{SIFT\xspace}
\newcommand{\gist}{GIST\xspace}

\newcommand\vldbdoi{XX.XX/XXX.XX}
\newcommand\vldbpages{XXX-XXX}
\newcommand\vldbvolume{14}
\newcommand\vldbissue{1}
\newcommand\vldbyear{2020}
\newcommand\vldbauthors{\authors}
\newcommand\vldbtitle{\systitle: Efficiently Supporting Conjunctions and Disjunctions for Filtered Approximate Nearest Neighbor Search} 
\newcommand\vldbavailabilityurl{URL_TO_YOUR_ARTIFACTS}
\newcommand\vldbpagestyle{plain} 

\begin{document}
\title{\systitle: Efficiently Supporting Conjunctions and Disjunctions \\ 
for Filtered Approximate Nearest Neighbor Search}

\author{Tianming Wu}
\affiliation{%
  \institution{UT Austin}
}
\email{tianming.wu@utexas.edu }

\author{Dixin Tang}
\affiliation{%
  \institution{UT Austin}
}
\email{dixin@utexas.edu}

\begin{abstract}
Filtered approximate nearest neighbor search (ANNS) restricts the search to data objects whose attributes satisfy a given filter and retrieves the top-$K$ objects that are most semantically similar to the query object.
Many graph-based ANNS indexes are proposed to enable efficient filtered ANNS
but remain limited in applicability or performance: 
indexes optimized for a specific attribute achieve high efficiency 
for filters on that attribute but fail to support complex filters
with arbitrary conjunctions and disjunctions over multiple attributes. 
Inspired by the design of relational databases,
this paper presents \sys,
a new indexing framework that allows users to selectively create 
ANNS indexes optimized for filters on specific attributes 
and employs a cost-based optimizer to efficiently utilize them
for processing complex filters. 
\sys includes three novel techniques: 
1) a new optimization metric that captures the tradeoff
between query execution time and accuracy, 
2) a two-phase optimization for handling filters with conjunctions and disjunctions, and 3) an index borrowing optimization that uses 
an attribute-specific index to process filters on another attribute. 
Experiments on four real-world datasets show that \sys
outperforms the best baseline by up to 9.8$\times$ 
in query throughput at recall 0.95. 
\end{abstract}

\maketitle

\pagestyle{\vldbpagestyle}
\begingroup\small\noindent\raggedright\textbf{PVLDB Reference Format:}\\
\vldbauthors. \vldbtitle. PVLDB, \vldbvolume(\vldbissue): \vldbpages, \vldbyear.\\
\href{https://doi.org/\vldbdoi}{doi:\vldbdoi}
\endgroup
\begingroup
\renewcommand\thefootnote{}\footnote{\noindent
This work is licensed under the Creative Commons BY-NC-ND 4.0 International License. Visit \url{https://creativecommons.org/licenses/by-nc-nd/4.0/} to view a copy of this license. For any use beyond those covered by this license, obtain permission by emailing \href{mailto:info@vldb.org}{info@vldb.org}. Copyright is held by the owner/author(s). Publication rights licensed to the VLDB Endowment. \\
\raggedright Proceedings of the VLDB Endowment, Vol. \vldbvolume, No. \vldbissue\ %
ISSN 2150-8097. \\
\href{https://doi.org/\vldbdoi}{doi:\vldbdoi} \\
}\addtocounter{footnote}{-1}\endgroup

\ifdefempty{\vldbavailabilityurl}{}{
\vspace{.3cm}
\begingroup\small\noindent\raggedright\textbf{PVLDB Artifact Availability:}\\
The source code, data, and/or other artifacts have been made available at \url{\vldbavailabilityurl}.
\endgroup
}

\input{intro}
\input{background}
\input{system}
\input{implementation}
\input{experiments}
\input{related}

\section{Conclusion}
This paper introduces \sys, an indexing framework 
designed for filtered ANNS. 
\sys allows DBMS administrators to create attribute-specific ANNS indexes on selective attributes and adopts  
a cost-based optimization framework to effectively utilize these indexes for improving the performance of filtered ANNS. 
The efficiency of \sys stems from three key innovations: 
the optimization metric for quantifying the tradeoff between search time and accuracy, 
the optimization algorithms for processing filters with conjunctions and disjunctions, and 
the index borrowing technique that enables leveraging one attribue-specific index to process filters on correlated attributes. 
Extensive experiments show that \sys significantly improves the performance of filtered ANNS with conjunctive and disjunctive filters.



\bibliographystyle{ACM-Reference-Format}
\bibliography{vectordb}

\end{document}

%% file: intro.tex
\section{Introduction}
Vector databases are the foundational infrastructure 
for \emph{semantic search} and have been adopted to support 
a broad range of information systems, 
such as retrieval-augmented generation systems for large language models~\cite{Chameleon:jiang2023chameleon,RAG:journals/corr/abs-2002-08909,RAGO:jiang2025rago,COT-RAG:conf/acl/TrivediBKS23}, recommendation systems~\cite{Analyticdb:wei2020analyticdb, recommendrajput2023recommender}, search engines~\cite{Lucence,OpenSearch,Elastic}, and knowledge bases~\cite{HQI:10.1145/3589777,GraphRAG:journals/corr/abs-2404-16130,SAGA:conf/sigmod/IlyasRKPQS22}. 
Vector databases support semantic search 
by encoding each data object, such as a document or an image, into a high-dimensional vector, and quickly 
but approximately finding the top-$K$ data objects that are most 
semantically similar to a query object (i.e., \emph{a similarity query}) based on vector distances~\cite{milvus,Milvus:wang2021milvus,chromadb,weaviate, lancedb,Analyticdb:wei2020analyticdb,qdrant,Single-V:pvldb/ChenJZPWWHSWW24, NSG:journals/pvldb/FuXWC19, HNSW:journals/pami/MalkovY20, DiskANN:conf/nips/SubramanyaDSKK19}, known as \emph{approximate nearest neighbor search (ANNS)}. 
When processing a similarity query with a filter 
on the attributes of 
the data objects (e.g., searching for the papers in 
the ``DB'' field and published after 2025), 
ANNS restricts the search to the 
subset of data objects passing the filter, 
known as \emph{filtered ANNS}. 

Efficiently and accurately supporting filtered ANNS remains challenging, as the performance of existing \emph{ANNS indexes} degrades significantly under complex filters. \emph{Graph-based indexes}, for example, are widely adopted to support ANNS 
due to their strong tradeoff between query execution time and accuracy~\cite{DiskANN:conf/nips/SubramanyaDSKK19,HNSW:journals/pami/MalkovY20,NSG:journals/pvldb/FuXWC19,NSW:journals/is/MalkovPLK14,RWalks-SIGMOD25}. 
The core of graph-based indexes is the \emph{\pgraph},
which represents data objects as vertices and connects each vertex to a bounded number of nearby vertices based on vector distances~\cite{DiskANN:conf/nips/SubramanyaDSKK19,HNSW:journals/pami/MalkovY20,NSW:journals/is/MalkovPLK14}.
However, the filter associated with a similarity query 
can induce a \emph{sparse or even disconnected subgraph}~\cite{FreshDiskANN:journals/corr/abs-2105-09613}, significantly degrading search efficiency and accuracy.


\begin{figure}[t]
    \centering
    \includegraphics[width=65mm]{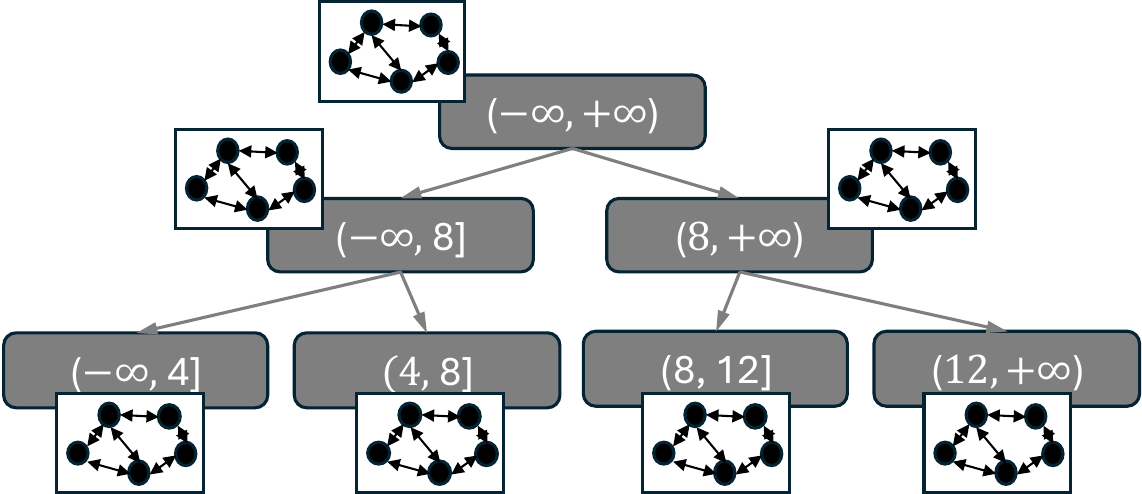}
    \vspace{-0mm}
    \caption{A tree-based graph index built on a numeric attribute. The attribute range is recursively partitioned, and for each tree node, a \pgraph is built over the data objects whose attribute values fall within the node's range.}
    \label{fig:tree-index}
    \vspace{-2mm}
\end{figure}

Recent studies have proposed new graph-based indexes to address the challenge of sparse graphs in filtered ANNS. However, these approaches remain limited in their applicability or performance. A line of research focuses on indexes optimized for filters on a single attribute of a particular data type~\cite{SeRF:10.1145/3639324,iRangeGraph:10.1145/3698814,NHQ:10.5555/3666122.3666814,Filtered-DiskANN:10.1145/3543507.3583552,WindowFilter:10.5555/3692070.3692567,CAPS:gupta2023capspracticalpartitionindex,UNIFY-VLDB25,RWalks-SIGMOD25,UNG:10.1145/3698822}, referred to as \emph{attribute-specific indexes}. Their goal is to construct sufficiently dense \pgraph{s} to process filters on the attribute they are optimized for. As a result, when processing filters on the attributes they are optimized for, they offer substantial performance advantages over graph-based 
indexes that support general filters~\cite{ACORN:journals/pacmmod/PatelKGZ24,HNSW:journals/pami/MalkovY20,DiskANN:conf/nips/SubramanyaDSKK19}. However, it remains an open challenge to effectively utilize attribute-specific indexes to support filters with arbitrary conjunctions and disjunctions.



\stitle{Our approach} 
This paper presents \sys, a novel graph-based indexing framework 
for efficient filtered ANNS that supports general filters 
with conjunctions and disjunctions.  
The design of \sys is guided by a key principle drawn from 
the practical considerations of relational databases:  
\textit{since it is prohibitively expensive to build indexes for all possible 
attribute combinations, DBMSs allow administrators to create indexes 
on selected attributes and rely on a cost-based optimizer to 
leverage these indexes for efficiently processing general multi-attribute filters.} 
Following this insight, \sys allows the vector database administrators 
to create attribute-specific ANNS indexes on selected attributes and designs a cost-based optimizer 
to utilize the available indexes to efficiently process similarity queries with complex filters.  

Specifically, \sys models data objects as a database relation, where the columns represent the embedding vector and associated attributes, which may be numeric or categorical.
It supports tree-based and hash-based 
graph indexes optimized for individual attributes. 
Tree-based graph indexes have shown state-of-the-art search performance for single-attribute range filters~\cite{WindowFilter:10.5555/3692070.3692567, DIGRA, iRangeGraph:10.1145/3698814, WoW} and support efficient updates~\cite{DIGRA, WoW}. 
We adopt a multi-way tree structure~\cite{DIGRA} and 
include a new cost-based method for efficiently 
using this index structure to support both single-attribute 
and multi-attribute filters.  
Figure~\ref{fig:tree-index} shows an example of this tree-based graph index. 
It recursively partitions the value range of an attribute, 
like a \btree, and builds a \pgraph 
for each node over the data objects whose corresponding 
attribute values fall into this node's range. 
Processing a range predicate involves selecting one or more \pgraph{s} 
from the tree.  
For instance, to process the range predicate $6 \leq value \leq 8$ 
using the index in Figure~\ref{fig:tree-index}, 
the system may choose to search the \pgraph corresponding to the node $(4, 8]$, 
as this graph contains the highest proportion of nodes that satisfy the filter.
As a complement to tree-based indexes, we 
use hash-based indexes to 
support categorical data and point predicates (e.g., topic = ``DB'' or topic in [``DB'', ``CV'']). 
It builds a \pgraph for all data objects 
with the same categorical value. 


\stitle{Technical challenges} 
\sys adopts a query optimizer that selects a subset of \pgraph{s} 
from attribute-specific indexes to efficiently process a similarity query with a filter (i.e., a \emph{filtered similarity query}). 
Building such an optimizer requires overcoming two key challenges. 
First, we need an optimization metric that captures the tradeoff 
between query execution time and accuracy for filtered ANNS. 
A higher value of this metric should indicate a better 
tradeoff between the two factors. 
Intuitively, we might prefer dense \pgraph{s} 
(i.e., the ones containing more nodes satisfying the predicate).  
However, such a metric would favor many small \pgraph{s}, 
which in turn increases execution time. 
Consider the tree-based index in Figure~\ref{fig:tree-index}. 
For a predicate $1 \leq value \leq 8$, 
prioritizing dense graphs 
will choose $(-\infty, 4]$ and $(4, 8]$ 
although the graph for $(-\infty, 8]$ might be a better choice. 
This motivates the need for a new metric that balances the 
density and the number of \pgraph{s} involved. 
Second, executing a similarity query is fast, typically completing 
in sub-milliseconds to a few milliseconds, which leaves a small 
optimization budget. 
Meanwhile, the optimizer must consider many combinations of \pgraph{s} 
from attribute-specific indexes for a complex filter predicate.  
It is challenging to select the subset 
of \pgraph{s} that can efficiently process the query 
within the tight optimization time budget. 

\stitle{\sys optimizer} 
\sys addresses these challenges with two key techniques:  
(1) a new metric that can quantify the efficiency of executing a filtered similarity query on a set of \pgraph{s} and can be quickly estimated to determine the relative ordering without computing exact values; and (2) a two-phase optimization process that efficiently selects the \pgraph{s} for answering a filtered similarity query. 

Specifically, we design a new metric, \emph{\utility}, that balances the graph density for a filter against the number of \pgraph{s}. This metric favors subsets of graphs that achieve a higher overall density while reducing the total number of graphs used. Moreover, since the optimizer only needs to compare the relative efficiency of different subsets of graphs, the \utility\ is carefully formulated so that our estimation method only needs to compute the components that determine their relative ordering, without performing costly cardinality estimation (i.e., estimating the number of nodes in a graph passing a filter). 
To process a filtered similarity query, \sys converts the filter predicate into \emph{disjunctive normal form (DNF)}, where conjunctive clauses are connected by disjunctions, and processes conjunctions and disjunctions sequentially.  
For each conjunctive clause, \sys efficiently identifies up to two promising subsets of \pgraph{s} per attribute-specific index and selects the subset with the highest \utility across all indexes.  
For disjunctions, \sys groups the \pgraph{s} selected for all conjunctive clauses by index and removes duplicates.  
Within each group, \sys adopts a novel algorithm that exploits the tree-based index structure to identify common ancestor \pgraph{s} that can subsume and replace the graphs in the group, thereby further improving the \utility. 


\stitle{Optimization: index borrowing}
Users may issue filter predicates on attributes for which no attribute-specific indexes are available.
While \sys can still process such queries using the \pgraph that covers all data objects in the relation (e.g., the graph for the root node in Figure~\ref{fig:tree-index}), its performance will degrade when handling complex filters.
To address this, \sys introduces an index-borrowing optimization that leverages an index built on one attribute to process a filter on another.
The key insight is that when two attributes are correlated, we can synthesize a new predicate on one attribute for which the index 
is available based on the input predicate. 
\sys then uses this synthesized predicate to select 
\pgraph{s} from the corresponding index 
that can more efficiently process the filtered similarity query.

\stitle{Evaluation}
To evaluate the effectiveness of \sys, we compare \sys with 
five baselines that support multi-attribute filters 
on numeric and categorical data. 
We use four datasets and generate query workloads 
that include filters with conjunctions and disjunctions.  
Our experiments show that \sys has up to 
9.8$\times$ higher throughput at recall 0.95 than the best baseline.

\stitle{Research vision} 
\sys opens a new direction for supporting filtered ANNS in vector databases by drawing on successful practice from relational databases: 
adopting a cost-based optimizer to best utilize 
available attribute-specific indexes to provide efficient filtered ANNS.  
This framework can potentially support 
new attribute-specific indexes or new data types (e.g., label data) and predicates (e.g., regex-based string matching), which is left for future work.
Moreover, it introduces new research opportunities, 
such as index compression for reducing memory consumption 
and automatic index recommendation~\cite{AIMeetsAI:conf/sigmod/DingDM0CN19} for filtered ANNS. 

%% file: background.tex
\section{Background and Problem Statement}
\label{sec:bg}

\stitle{Vector databases and ANNS}
Vector databases support approximate nearest neighbor search (ANNS) by 
converting data objects (e.g., images or documents) into high-dimensional vectors 
and approximately retrieving the 
$K$ most semantically similar objects 
for a given query (i.e., \emph{similarity queries}) 
based on the distances between the vector of the query and the vectors stored in the database~\cite{milvus,Milvus:wang2021milvus,chromadb,weaviate, lancedb,Analyticdb:wei2020analyticdb, qdrant, Single-V:pvldb/ChenJZPWWHSWW24}. 
The accuracy is measured by recall:  
$recall\text{@}K=\frac{|R \cap R'|}{K}$, 
where $R$ is top-$K$ nearest neighbors 
returned by ANNS   
and $R'$ is the ground truth top-$K$ result.

\stitle{Graph-based indexes} 
Vector databases rely on indexes to perform ANNS.  
Graph indexes, such as HNSW~\cite{HNSW:journals/pami/MalkovY20} and Vamana~\cite{DiskANN:conf/nips/SubramanyaDSKK19}, have been widely adopted due to their strong performance in balancing search time and accuracy, particularly in high-dimensional vector spaces. 
The core of graph indexes is using \emph{\pgraph{s}} 
to guide similarity search. 
In a \pgraph, each data object is represented as a vertex and 
connects to a bounded number of nearby vertices via directed edges based on the vector distances. 



Finding the top-$K$ nearest neighbors adopts \emph{best-first search}~\cite{NSW:journals/is/MalkovPLK14,HNSW:journals/pami/MalkovY20,DiskANN:conf/nips/SubramanyaDSKK19,NSG:journals/pvldb/FuXWC19}. 
The algorithm maintains a bounded size of \emph{search queue} that stores candidate vertices for answering the similarity query. Starting from one or more entry points, the unexpanded vertex that is closest to the query vector in the queue is expanded by adding its neighbors to the queue, which only keeps a fixed number of vertices closest to the query vector. 
The search continues until convergence, typically when no newly expanded neighbors are closer than the farthest vertex in the queue. The size of the queue is configurable, allowing for a tradeoff between search time and accuracy. 

\begin{figure}[!t]
    \centering
    \includegraphics[width=70mm]{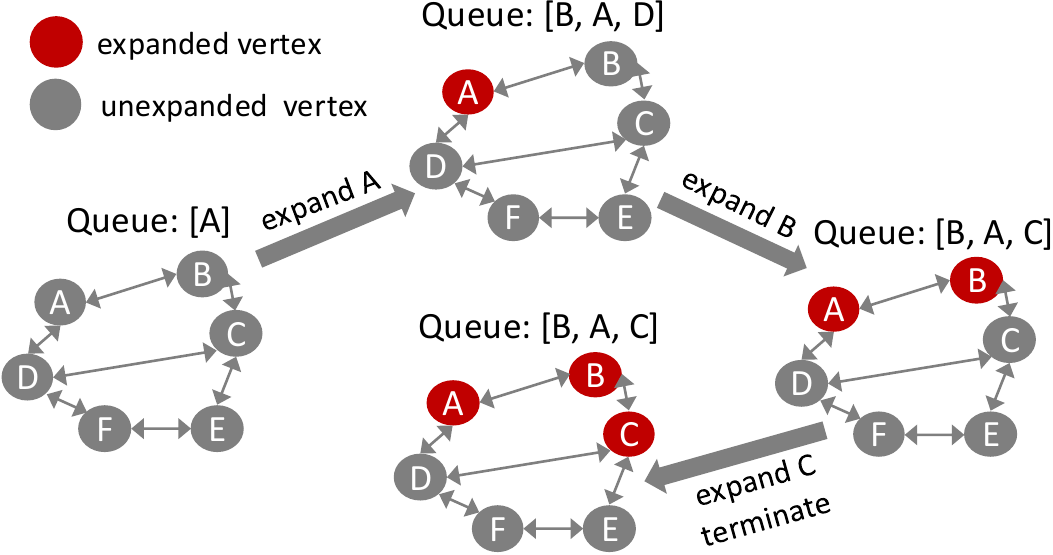}
    \vspace{-3mm}
    \caption{An example illustrating best-first search on a proximity graph for a graph index}
    \label{fig:graph_example}
    \vspace{-5mm}
\end{figure}

Figure~\ref{fig:graph_example} shows an example of best-first search. 
The search begins at vertex A, with the queue initially containing A. 
After expanding A, its vertices B and D are added to the queue. 
The algorithm then picks the vertex with 
the smallest distance to the query vector (i.e., B in this example), 
expands it, and updates the queue with its unvisited neighbor, C. 
Since the queue has a maximum size of 3, the farthest 
vertex to the query in the queue, D, is removed. 
Finally, the search terminates because C's unvisited neighbor E 
has a larger distance to q than the farthest node currently 
in the queue (i.e., C). 

HNSW~\cite{HNSW:journals/pami/MalkovY20} uses a hierarchy of \pgraph{s} to 
quickly locate an entry point in the bottom layer that is likely near the region that includes the nearest neighbors to the query vector. 
The Vamana graph~\cite{DiskANN:conf/nips/SubramanyaDSKK19} simplifies the design by using a single \pgraph without hierarchy and introduces long-range edges to accelerate convergence toward the region closest to the query vector. \sys uses the Vamana graph as its \pgraph. 


\stitle{Filtered ANNS} 
A similarity query often includes a filter on data object attributes, such as price, topic, or timestamp. The filter restricts the ANNS search to only those objects that satisfy the condition, a problem known as \emph{filtered ANNS}. 

The primary challenge for efficiently supporting filtered ANNS 
is that the filter induces a sparse or even disconnected \pgraph. 
Three basic strategies are adopted for supporting filtered ANNS. The \emph{pre-filtering} strategy skips the ANNS index by comparing the query vector with all vectors that satisfy the filter; it is only effective when the filter selectivity is extremely low~\cite{ACORN:journals/pacmmod/PatelKGZ24, Analyticdb:wei2020analyticdb}. 
The \emph{in-filtering} strategy applies filters during the similarity search on the ANNS index (e.g., a graph index).
The \emph{post-filtering} strategy first searches the index and then applies the filter to the retrieved results.  
If fewer than $K$ objects satisfy the filter, the search is retried with a longer search queue to find additional valid candidates.  

Beyond the basic strategies, recent studies have proposed new graph indexes to address the challenge of sparse graphs ~\cite{SeRF:10.1145/3639324,ACORN:journals/pacmmod/PatelKGZ24,iRangeGraph:10.1145/3698814,WindowFilter:10.5555/3692070.3692567,UNG:10.1145/3698822,DIGRA,UNIFY-VLDB25,SIEVE-VLDB25, HQI:10.1145/3589777, RWalks-SIGMOD25, WoW}. 
They materialize additional edges, construct filter-specific \pgraph{s}, or visit nodes that do not pass the filter (i.e., \emph{out-of-range} nodes) to ensure that the search is performed over a sufficiently dense graph. 
However, existing methods remain limited in applicability or performance. One line of work builds graph indexes tailored to a single attribute of a particular data type~\cite{SeRF:10.1145/3639324,WindowFilter:10.5555/3692070.3692567,UNIFY-VLDB25,iRangeGraph:10.1145/3698814,DIGRA,UNG:10.1145/3698822,RWalks-SIGMOD25, WoW, CAPS:gupta2023capspracticalpartitionindex, NHQ:10.5555/3666122.3666814}, such as constructing \pgraph{s} for different subranges of a numeric attribute. These approaches achieve significantly higher performance than general indexes that support arbitrary filters~\cite{ACORN:journals/pacmmod/PatelKGZ24, HNSW:journals/pami/MalkovY20, DiskANN:conf/nips/SubramanyaDSKK19}, but cannot efficiently support multi-attribute filters with arbitrary conjunctions and disjunctions. Other studies support complex filters but require that the filter workload is known~\cite{SIEVE-VLDB25,HQI:10.1145/3589777}.


\stitle{Problem statement}
We aim to build an indexing framework, \sys, which  
leverages high-performance attribute-specific indexes~\cite{iRangeGraph:10.1145/3698814,DIGRA, WindowFilter:10.5555/3692070.3692567} 
to support multi-attribute filters with 
conjunctions and disjunctions. 
This framework is similar to the access path selection 
framework in relational databases which best uses  
the indexes built on specific attributes to process 
multi-attribute filters.

\sys uses a relation $T(pk, a_1, a_2, \dots, a_k, \mathsf{object}, \mathsf{vector})$ to represent data objects, 
each of which is modeled as a tuple including a primary key $pk$, $k$ attributes $a_1, a_2, \dots, a_k$, a data object, 
and the vector embedding for the object.  
We assume a collection of attribute-specific graph-based indexes 
$I = \{I_{a_i} \mid a_i \in A_I\}$, 
where $A_I \subseteq \{a_1, a_2, \dots, a_k\}$ is the subset of attributes for which indexes are built, and each $I_{a_i}$ is an index on attribute $a_i$\footnote{Existing attribute-specific indexes are mainly designed for individual attributes; our framework can be naturally extended to multi-attribute indexes.}. 

The research problem \sys addresses is \emph{how to utilize the index collection $I$ to answer filtered similarity queries such that the system has the best tradeoff between query throughput and recall}.  

%% file: system.tex
\begin{figure*}[!t]
    \centering
    \includegraphics[width=150mm]{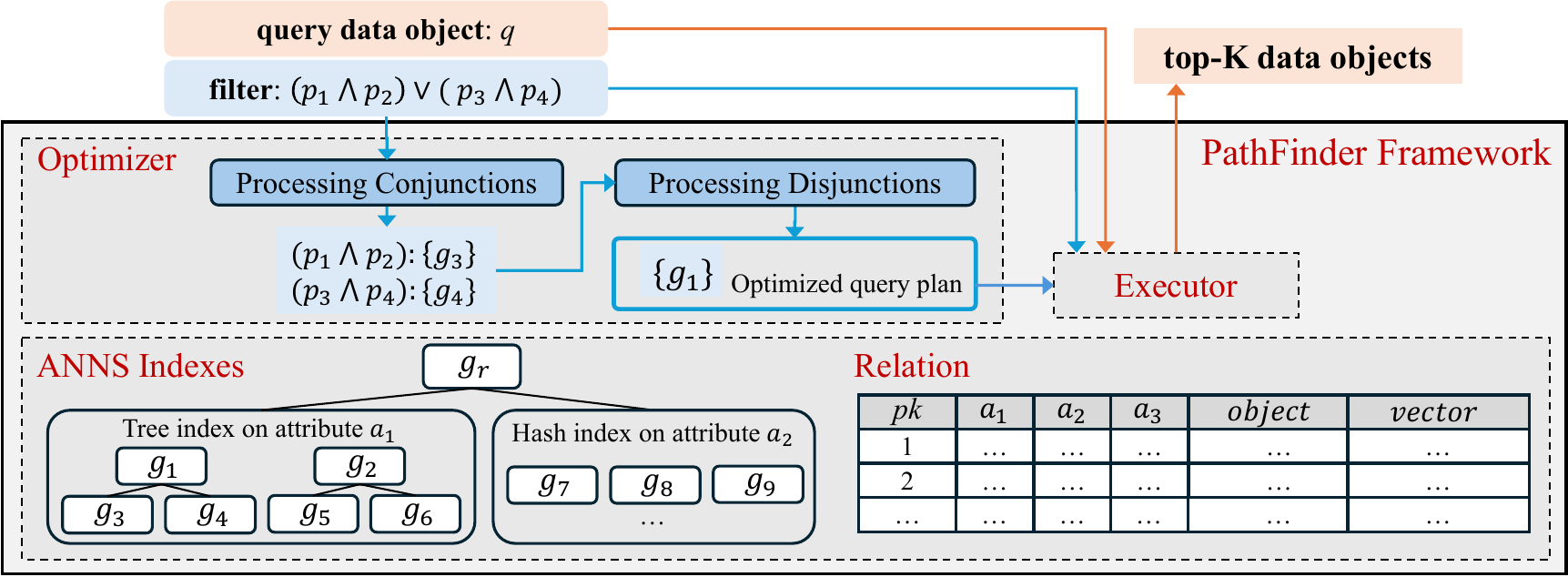}
    \vspace{-0mm}
    \caption{The workflow of \sys for processing an example query}
    \label{fig:overview}
    \vspace{-3mm}
\end{figure*} 

\section{System Designs} 
We now present the designs of \sys. 
We first give an overview of the framework 
and then describe the specific techniques in detail. 

\subsection{\sys Overview}
\label{sec:overview}

\sys represents a set of data objects along with their attributes and vectors as a relation 
$T(pk, a_1, a_2, \dots, a_k, \mathsf{object}, \mathsf{vector})$, and assumes DBMS administrators have chosen to create 
a collection of graph indexes $I$ for a subset of 
attributes, with each graph index $I_{a_i}$ 
corresponding to the attribute $a_i$. 
Each attribute can be either a numeric value or a categorical value represented as a string. 

\stitle{Supported indexes and predicates} 
\sys currently supports tree-based graph indexes~\cite{DIGRA} for 
range and point filters and 
hash-based graph indexes for point filters only. 
Specifically, the tree-based graph index adopts a 
multi-way tree structure~\cite{DIGRA} 
that recursively partitions the value range of an attribute, similar to a \btree, and builds a \pgraph for each tree node. 
Figure~\ref{fig:tree-index} illustrates an example, where 
the value range of a non-leaf node is partitioned into two. 
A hash-based graph index supports categorical attributes 
without ordering. 
It employs a hash table to map each categorical value to the partition of tuples with that value, and then builds a \pgraph for each partition. 
For example, building a hash-based graph index for the ``topic'' attribute of a set of research papers will partition the papers by topic (e.g., ``DB'' vs. ``CV'') and construct a separate \pgraph for each partition. 
We choose the Vamana graph~\cite{DiskANN:conf/nips/SubramanyaDSKK19} as the \pgraph in \sys. 

Figure~\ref{fig:overview} shows an example of a tree-based index 
and a hash-based index built on attributes $a_1$ and $a_2$, respectively. 
\sys builds a \pgraph for all tuples by default (e.g., $g_r$ in Figure~\ref{fig:overview}), 
which also serves as the root node for 
both the tree-based and hash-based indexes. 
That is, $g_r$ is the parent node of the \pgraph{s} 
$g_1$ and $g_2$ for the tree-based index 
and is the parent node of $g_7$-$g_9$ for the hash-based index. 
Therefore, the hash-based index trivially adopts a two-layer tree structure. 

A tree-based graph index for an attribute $a_i$ 
supports a variety of range predicates, such as $a_i \geq c$, $a_i \leq c$, $a_i = c$, $a_i < c$, and $a_i > c$, where $c$ is a literal. 
A hash-based graph index supports categorical predicates, 
including equality ($a_i = c$) and membership ($a_i \text{ IN } S$ for a set of values $S$). 
\sys supports Boolean combinations of 
these atomic predicates through conjunctions and disjunctions.

\stitle{\sys optimizer} 
Given a filtered similarity query and a set of attribute-specific indexes that comprise \pgraph{s}, 
\sys adopts a cost-based approach that 
selects a subset of \pgraph{s} to best process this query. 
To guide the optimization, we define a novel optimization metric,
\emph{\utility}, which jointly balances recall and search time  
to favor subsets of graphs that achieve higher overall density with respect to the filter predicate while reducing the number of graphs to search. 
Moreover, the \utility can be quickly estimated to decide the relative orderings without computing the exact values.
We present the definition of \utility and the estimation method in Section~\ref{sec:utility} 

\sys then includes a two-phase optimization mechanism that quickly finds an 
execution plan for processing the filtered similarity query 
while maximizing the \utility. 
Specifically, the execution plan is represented as the subsets of \pgraph{s} 
selected from each available index $I_{a_i}$, referred to as a \emph{\gsp}. 
\sys then searches each \pgraph in this plan and combines their results 
to return the top-K data objects to users. 

Figure~\ref{fig:overview} shows the workflow for processing an example filtered similarity query. 
\sys processes the filter predicate $p$ 
in disjunctive normal form (DNF), 
which expresses $p$ as a disjunction of conjunctive clauses: 
$p = C_1 \lor C_2 \lor \cdots \lor C_m$, 
where each clause $C_i$ is a conjunction of atomic predicates 
on individual attributes. 

For a conjunctive clause $C_i$, the optimizer considers all indexes involved in $C_i$, finds up to two promising \gsp{s} for each index, 
and chooses the one with the highest \utility across all indexes. 
For the example in Figure~\ref{fig:overview}, 
we have $C_1 = p_1 \land p_2$ and $C_2 = p_3 \land p_4$. 
\sys selects $\{g_3\}$ and $\{g_4\}$ for $C_1$ 
and $C_2$, respectively. 
To process disjunctions, a \naive approach is to execute the \gsp for each conjunctive clause independently.
\sys improves upon this by combining the \gsp{s} 
for more efficient execution.
Specifically, \sys merges and deduplicates the plans, groups their proximity graphs by index, and leverages the index hierarchy to identify ancestor graphs that can replace descendant graphs 
to reduce redundancy and improve execution efficiency. 
For example, in Figure~\ref{fig:overview},
given $\{g_3, g_4\}$ as the output from the previous phase,
\sys may select $\{g_1\}$ to replace $\{g_3, g_4\}$ because 
$g_1$ covers the value ranges of $g_3$ and $g_4$ and may be 
more efficient to search (depending their relative \utility values). 
We describe processing conjunctions and disjunctions in Sections~\ref{sec:conjunction}--\ref{sec:disjunction}. 
We include an optimization in Section~\ref{sec:optimization}, which leverages existing indexes to process predicates on attributes that lack dedicated indexes.


\subsection{Search Utility}
\label{sec:utility} 
\sys selects a subset of \pgraph{s} from the available indexes 
to efficiently answer a filtered similarity query. 
Intuitively, \pgraph{s} that have a higher fraction of nodes 
passing the filter (i.e., dense graphs) are preferred, 
as searching them improves recall and reduces search time. 
However, using too many dense graphs, such as all leaf nodes covered by a predicate in a tree-based index, adds a linear factor to the otherwise logarithmic graph search complexity, increasing overall search time. 
\sys therefore introduces an optimization metric that balances 
the two factors. 

Designing such a metric presents a key challenge: 
minimizing the time cost of evaluating its value. 
Executing a similarity query is fast, 
typically within sub-milliseconds to a few milliseconds, 
leaving a tight time budget for the optimizer. 
Estimating graph density (i.e., estimating the fraction of graph nodes satisfying the predicate) 
requires cardinality estimation, which takes 
non-trivial time for a complex filter. 
For example, a recent histogram-based estimator~\cite{FLAT:journals/pvldb/ZhuWHZPQZC21} takes approximately 0.2--0.5ms to perform a single cardinality estimation for a filter involving multiple attributes, which introduces a substantial overhead to the execution time of a similarity query.

To address this challenge, we adopt the following key observation:
the optimizer only needs to \emph{rank} different execution plans according to the optimization metric, rather than compute their exact values. 
Therefore, we design the metric to capture the effects of both graph density and the number of graphs, while structuring it so that only part of the metric can be quickly estimated to determine the relative ordering among different plans. 

\stitle{Search utility definition} 
We define the optimization metric on a set of \pgraph{s} that have disjoint nodes because this requirement simplifies both the definition and estimation of the metric. The metric under this requirement is sufficient for our optimization framework. Formally, 
we define \emph{\utility} $U(G, p)$ 
to represent the efficiency of using a set of 
disjoint \pgraph{s} $G$ to process 
a similarity query with a filter predicate $p$: 
\begin{equation}
U(G, p) =
\begin{cases}
\displaystyle
\frac{\text{card}(R, p)}
{\sum_{g_i \in G} \text{card}(g_i) \times |G|^{\alpha}}, & \text{if $G$ covers $p$}, \\[10pt]
0, & \text{otherwise.}
\end{cases}
\label{eq:utility}
\end{equation}
Here, $R$ represents the relation storing all tuples. 
The requirement that $G$ covers $p$ means 
$G$ must include all tuples in relation $R$ that satisfy $p$; otherwise, the utility is zero. 
The term $\text{card}(R, p)$ represents the total number 
of tuples passing $p$, 
and $\sum_{g_i \in G} \text{card}(g_i)$ 
denotes the total number of tuples in $G$. 
Thus, their ratio, $\frac{\text{card}(R, p)}{\sum_{g_i \in G} \text{card}(g_i)}$, 
captures the overall density of $G$ with respect to $p$. 
The factor $|G|^{\alpha}$ penalizes the use of a larger number of graphs, 
where $\alpha$ controls the intensity of this penalty. 
Our experiments show that setting $\alpha = 0.4$ yields the best performance. 
The value of $U(G, p)$ is in $[0, 1]$, where a higher value 
indicates higher search efficiency.

\stitle{Estimation method} 
Estimating the exact value of $U(G, p)$ may be time-consuming 
because it requires computing the cardinality $\text{card}(R, p)$. 
Fortunately, $\text{card}(R, p)$ remains constant across different $G$s 
for the same predicate $p$. 
Therefore, to compare the utilities of different plans, 
it suffices to evaluate 
$\sum_{g_i \in G} \text{card}(g_i) \times |G|^{\alpha}$ 
for ranking purposes. 
Computing $\text{card}(g_i)$ is efficient, as the number of tuples 
in each \pgraph can be precomputed.


\begin{figure}[!t]
    \centering
    \includegraphics[width=60mm]{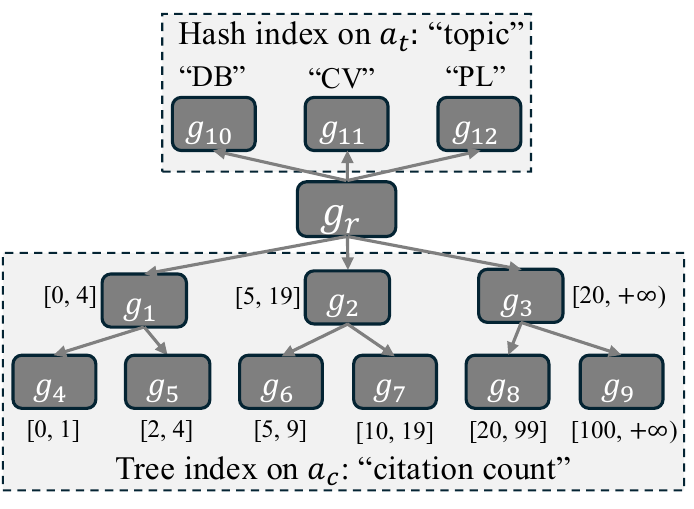}
    \vspace{-3mm}
    \caption{A tree-based graph index and a hash-based graph index built on the ``citation count'' and ``topic'' attributes.}
    \label{fig:conjunction_example}
    \vspace{-5mm}
\end{figure}

\subsection{Processing Conjunctions}
\label{sec:conjunction} 
Given a filter predicate in DNF  
$p = C_1 \lor C_2 \lor \cdots \lor C_m$, 
\sys first identifies a \gsp (i.e., a subset of \pgraph{s}) for each conjunctive clause $C_i$ 
that has the highest \utility. 
Since enumerating the exponential number of possible \gsp{s} is prohibitively expensive, \sys employs a greedy algorithm to reduce the optimization time while still finding a high-quality \gsp. For this and the next subsection, we only consider indexes on the attributes involved by $C_i$. We will discuss relaxing this assumption in Section~\ref{sec:optimization}. 


\stitle{Key ideas}
\sys identifies up to two high-quality \gsp{s} that can process $C_i$ for each index and selects the one with the highest $U(G, C_i)$ across all indexes. 
The two plans are complementary: the first includes a single \pgraph, while the second consists of multiple smaller and denser graphs. 
\sys efficiently finds these two plans by leveraging the monotonicity property of $U(G, C_i)$: for two \pgraph{s} $g_i$ and $g_j$ where $g_j$ is a child of $g_i$ and both cover $C_i$, we have $U(\{g_i\}, C_i) > U(\{g_j\}, C_i)$ because $\text{card}(g_j)$ is smaller than $\text{card}(g_i)$.

Based on this property, 
for the first plan, \sys 
starts with the root node $g_r$ 
and recursively selects a child node to replace its parent 
until the child no longer covers $C_i$ 
or a leaf node is reached.
Figure~\ref{fig:conjunction_example} shows an example of two indexes. If the predicate is:   
$(2 \leq a_c \leq 10) \land a_t = \text{``DB''}$, 
\sys selects $\{g_r\}$ and $\{g_{10}\}$  
as the first \gsp{s} for the tree index 
and hash index, respectively.  
This process yields a \gsp consisting of a single \pgraph (denoted $g_s$) for each index. 

If $g_s$ has child nodes, \sys further constructs the 
second \gsp using its descendant nodes. 
This second plan complements the first 
by combining multiple smaller, denser graphs to process $C_i$. 
Specifically, \sys partitions $C_i$ based on the predicates of the child nodes of $g_s$ 
and selects up to one \pgraph to process each partition. 
Formally, $C_i = (C_i \land p_{s1}) \lor \cdots \lor (C_i \land p_{sk})$, where $p_{sj}$ represents the predicate associated with the $j$th child of $g_s$. 
For each child node whose predicate $p_{sj}$ overlaps with $C_i$, \sys finds the \pgraph 
with the highest \utility within the subtree rooted at this child node, 
again leveraging the monotonicity property. 
Finally, the second plan consists of all \pgraph{s} 
selected for all partitions of $C_i$. 
The second plan, therefore, contains at most as many \pgraph{s} as the fan-out factor of the tree structure.

Consider the earlier predicate example $(2 \leq a_c \leq 10) \land a_t = \text{``DB''}$ 
for the indexes in Figure~\ref{fig:conjunction_example}. 
For the hash index, no second plan is generated. 
For the tree-based index, the first plan is $\{g_r\}$ 
and $g_r$ has three child nodes, 
two of which overlap with the input predicate 
(i.e., $g_1$ and $g_2$). 
The partition of the input predicate for $g_1$ is: 
$(2 \leq a_c \leq 10) \land (0 \leq a_c \leq 4) \land a_t = \text{``DB''}$, which can be simplified as: $(2 \leq a_c \leq 4) \land a_t = \text{``DB''}$. Using this predicate to 
search the subtree of $g_1$, we select $g_5$ as it is a leaf node covering this predicate. 
Similarly, we select $g_2$ for the subtree rooted at $g_2$. 
So the second plan for the tree-based index is 
$\{g_5, g_2\}$. 
Finally, \sys combines candidate plans from both indexes 
and chooses the one with the highest $U(G, C_i)$ 
from the candidate set $\{\{g_{10}\}, \{g_r\}, \{g_2, g_5\}\}$. 

\begin{algorithm}[t]
\caption{Processing a conjunctive clause}
\label{alg:conjunction}
\small
\DontPrintSemicolon
\KwIn{
$C_i$: a conjunctive clause, $I$: indexes involved by $C_i$
}
\KwOut{ A \gsp for $C_i$}
\SetKwFunction{FindSingleGraph}{FindSingleGraph}
\SetKwProg{Fn}{Function}{:}{}
\SetKwFunction{FindSecondPlan}{FindSecondPlan}

    $\mathcal{G} \leftarrow \emptyset$\;
    \ForEach{$I_k$ in $I$}{
        $g_r \leftarrow$ the root node of $I_k$\; 
        $g_s \leftarrow$ \FindSingleGraph$(g_r, C_i)$\;
        $\mathcal{G} \leftarrow$ add $\{ g_s \}$ to $\mathcal{G}$\;
        \If{$g_s$ has child nodes}{
            $G \leftarrow$ \FindSecondPlan$(g_s, C_i)$\;
            $\mathcal{G} \leftarrow$ add $G$ to $\mathcal{G}$\;
        }
    }
\Return $\arg\max_{G \in \mathcal{G}} U(G, C_i)$\;

\BlankLine
\Fn{\FindSingleGraph($g$, $p$)}{
    \If{$g$ is leaf}{
        \Return $g$\;
    }
    \ForEach{$g_c$ in $g$'s child nodes}{
        \If{$g_c$ covers $p$}{
            \Return \FindSingleGraph($g_c$, $p$)\;
        }
    }
    \Return $g$\;
}

\BlankLine
\Fn{\FindSecondPlan($g_s$, $p$)}{
    $G \leftarrow \emptyset$\;
    \ForEach{$g_c$ in $g_s$'s child nodes}{
        $p_{sc} \leftarrow g_c\text{'s predicate}$\;
        \If{$p$ overlaps with $p_{sc}$}{
            $g_c^* \leftarrow$ \FindSingleGraph($g_c$, $p \land p_{sc}$)\;
            $G \leftarrow G \cup \{ g_c^* \}$\;
        }
    }
    \Return $G$\;
}
\end{algorithm}

\stitle{Algorithm description}
Algorithm~\ref{alg:conjunction} shows how \sys selects the \gsp 
for a conjunctive clause $C_i$. 
Given the set of indexes $I$ involved by $C_i$, \sys enumerates each index $I_k$
to identify two candidate plans. It starts from the root node $g_r$ and calls 
\FindSingleGraph to find the deepest graph  $g_s$ that still covers $C_i$,
leveraging the monotonicity of the utility function $U(G, C_i)$. 
If $g_s$ has child nodes, \sys invokes \FindSecondPlan 
to construct a complementary plan using multiple smaller \pgraph{s}.
This function partitions $C_i$ by the predicates of the children of $g_s$ and finds the best \pgraph in each child's subtree.   
For all candidate plans from all indexes, 
\sys selects the one with the highest \utility. 
The worst case of this algorithm will visit all $N$ nodes of all $M$  
indexes, with the complexity of $O(N\times M)$. 

\begin{figure}[!t]
    \centering
    \includegraphics[width=60mm]{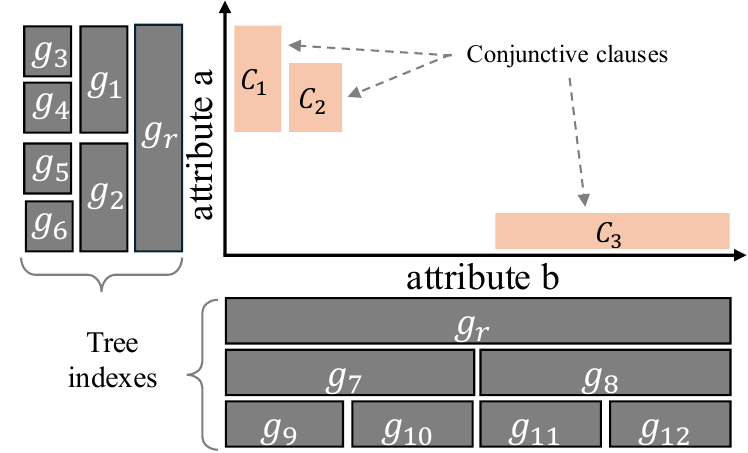}
    \caption{A predicate with three conjunctive clauses on two attributes; tree-based indexes are built for both attributes.}
    \label{fig:disjunction_example}
\end{figure}

\subsection{Processing Disjunctions}
\label{sec:disjunction}
The first phase of \sys's optimization selects a \gsp for 
each conjunctive clause $C_i$ of the filter predicate 
$p = C_1 \lor C_2 \lor \cdots \lor C_m$. 
To process disjunctions, one na\"ive method is to execute 
each plan one by one and combine their results. 
\sys, instead, considers merging these plans 
to optimize them. 

\stitle{Key ideas} 
The conjunctive clauses of a filter predicate may 
define value regions that are close to or 
overlap with each other. 
In such cases, the \gsp{s} selected for different conjunctive clauses 
may include \pgraph{s} that are duplicated or cover 
adjacent or overlapping regions.  
It is beneficial to 
deduplicate \pgraph{s} or replace smaller graphs 
with a larger one to further improve \utility.  

Figure~\ref{fig:disjunction_example} illustrates this idea 
using an example filter predicate 
$p = C_1 \lor C_2 \lor C_3$ on two attributes, $a$ and $b$. 
Each clause $C_i$ corresponds to a rectangular value region. 
Assuming \sys selects $\{g_9\}$ to process both $C_1$ and $C_2$, 
we can merge the two plans to remove a redundant $g_9$. 
In addition, if \sys selects $\{g_8\}$ to process $C_3$, it might 
be beneficial to use $\{g_r\}$ to replace $\{g_8, g_9\}$ 
to process all three conjunctive clauses, depending on 
the relative \utility of the two plans. 

One thing to note is that for a set of \pgraph{s} $G$ belonging to the same index, we do not simply remove the \pgraph{s} in $G$ whose ancestor graph is also in $G$. 
This is because the ancestor may be sparse with respect to the input filter and searching only this sparse graph while removing smaller, denser graphs can harm recall, as verified by our experiments (omitted due to space limits).
Instead, we consider replacing a subset of \emph{disjoint} \pgraph{s} with their ancestor, allowing \sys to use the \utility metric to decide. Although this optimization is not optimal, it avoids costly cardinality estimation and keeps the optimization time small.


\begin{algorithm}[t]
\caption{Optimizing a set of graphs for an index}
\label{alg:disjunction}
\KwIn{$G$: the set of selected graphs for an index $I$; $p$: the filter predicate}
\KwOut{Optimized graph set}
\small

\ForEach{leaf node $g$ in $I$}{
$g.\text{plan} \leftarrow (g \in G) \; ? \; \{g\} \; : \; \emptyset$\;
}

\ForEach{non-leaf node $g_p$ in bottom-up order in $I$}{
    $S \leftarrow \emptyset$\;
    \ForEach{child $g_c$ of $g_p$}{
        $S \leftarrow S \cup g_c.\text{plan}$\;
    }
    $p_{p} \leftarrow g_p\text{'s predicate}$\;
    \If{$U(S, p \land p_p) < U(\{g_p\}, p \land p_p)$}{
		$g_p.\text{plan} \leftarrow \{g_p\}$\;
		$G \leftarrow (G\setminus S) \cup \{g_p\}$\;
    }
	\Else{
		\If{$g_p \notin G$}{
			$g_p.\text{plan} \leftarrow S$\;
		}
        \Else{
			$g_p.\text{plan} \leftarrow \{g_p\}$\;
        }
    }
}
\Return{$G$}
\end{algorithm}

\stitle{Algorithm description} 
Given the \pgraph{s} from the \gsp{s} of all conjunctive clauses, \sys deduplicates the graphs, groups them by index, 
and optimizes each group independently.
Let $G$ denote the set of graphs selected from one index.
\sys employs an optimization algorithm that iterates through non-leaf nodes from bottom-up and, for each node $g_p$, replaces the disjoint descendant graphs of $g_p$ in $G$ with $g_p$ itself if $g_p$ has a higher \utility, 
as shown in Algorithm~\ref{alg:disjunction}. 
Specifically, each node $g$ maintains a variable $g.\text{plan}$ that records a set of disjoint graphs in $G$ 
and $g$'s subtree. 
This set of disjoint graphs serves as a candidate that 
may later be replaced by an ancestor graph. 
For a leaf node, $g.\text{plan}$ is initialized to ${g}$ if $g \in G$, and to $\emptyset$ otherwise.
For each non-leaf node $g_p$, the algorithm collects the disjoint descendant graphs that could be replaced by ${g_p}$ (i.e., $S$) and compares their relative \utility values to 
decide whether to replace.
If so, $g_p.\text{plan}$ is updated to $\{g_p\}$ 
and $G$ is updated accordingly (Lines 9–10).
Otherwise, if $g_p \notin G$, we retain $S$ as the 
disjoint graph set to be replaced later.
If $g_p \in G$, we instead start a new disjoint graph set by setting $g_p.\text{plan}$ to $\{g_p\}$ since $g_p$ overlaps 
with the graphs in $S$. 
In the worst case, the algorithm visits every node in every index, resulting in the same $O(N \times M)$ complexity as Algorithm~\ref{alg:conjunction}.




\subsection{Index Borrowing}
\label{sec:optimization}
The previous two subsections assume that, for a similarity query with a filter predicate $p$, \sys only leverages indexes built on the attributes involved in $p$. 
However, when two attributes $a$ and $b$ are correlated, 
it is beneficial to utilize an index built on attribute $a$ to process the filter predicate on the correlated attribute $b$ 
if no index is built on attribute $b$, rather than \naive{ly} searching the root node $g_r$. 
For example, assume all tuples satisfying $b \leq 5$ also satisfy $a \leq 6$. To process $b \leq 5$, we can use the predicate $a \leq 6$ to find a \gsp from the index built on $a$, and then apply the filter $b \leq 5$ during query execution to obtain the top-$K$ results. 



\stitle{Key ideas and algorithm} 
Assume that we want to use the index $I_a$ for attribute $a$ to process a predicate $p_b$ on attribute $b$. 
Our key ideas are: 
(1) synthesizing a predicate $p_a$ on attribute $a$ such that the tuples satisfying $p_b$ are a subset of those satisfying $p_a$ (i.e., $p_a$ \emph{covers} $p_b$) while minimizing the number of tuples passing $p_a$; and 
(2) using the synthesized $p_a$ to find a \gsp in $I_a$ via Algorithm~\ref{alg:conjunction}, and apply $p_b$ during query execution. 
We require that $p_a$ covers $p_b$ to ensure the \gsp selected for $p_a$ includes all tuples defined by $p_b$.

To quickly synthesize $p_a$ from $p_b$,
\sys precomputes, for each \pgraph in $I_a$, the value range of attribute $b$ among its tuples.
In a hash-based index $I_a$, each \pgraph corresponds to a categorical value of attribute $a$.
For each graph, \sys checks whether its value range for attribute $b$ (denoted as $g.p_b$) overlaps with $p_b$.
If so, the corresponding categorical value of $a$ 
is added to a set $C$. 
Finally, \sys synthesizes the predicate $p_a$ as ``$a \text{ IN } C$''. 
For a tree-based index, we aim to synthesize a range predicate $p_a$ from $p_b$. 
The key idea is to determine the minimum and maximum boundaries for $p_a$ by scanning the leaf nodes of the tree-based index and checking for overlap with $p_b$. 
Specifically, to find the minimum boundary, we scan the leaf nodes from left to right and, for each \pgraph $g$, check whether $g.p_b$ overlaps with $p_b$. 
This process stops at the first $g$ that overlaps with $p_b$ 
and the minimum value of $g.p_a$ is then used as the lower boundary of $p_a$. 
Similarly, we scan the leaf nodes from right to left to determine the maximum value of $p_a$. 

\begin{figure}[!t]
    \centering
    \includegraphics[width=80mm]{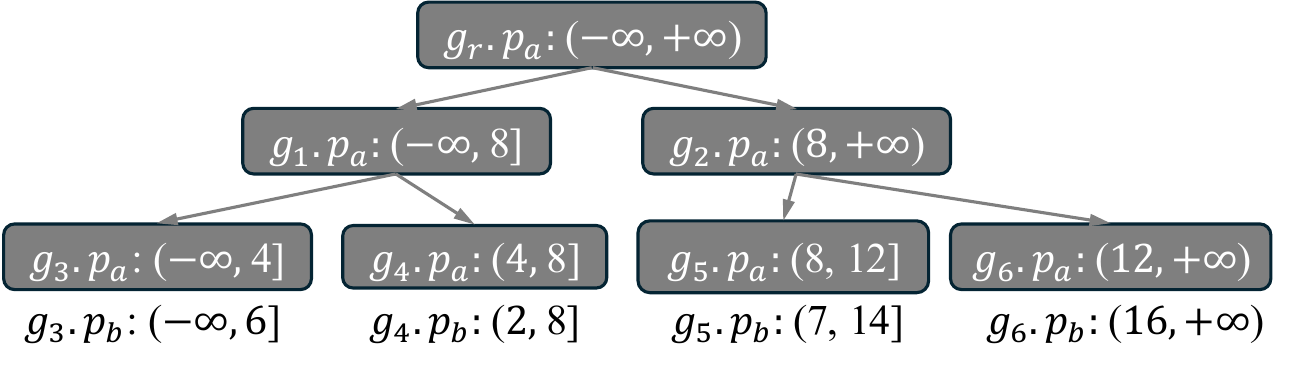}
    \vspace{-5mm}
    \caption{A tree-based index built on attribute $a$. It includes attribute $b$'s value ranges (i.e., $p_b$) for the leaf nodes.}
    \label{fig:optimization}
    \vspace{-5mm}
\end{figure}

\stitle{Example} 
Figure~\ref{fig:optimization} shows an example of a tree-based index built on attribute $a$.
Each node records the value ranges of attributes $a$ and $b$ that its tuples fall into, 
denoted as $g.p_a$ and $g.p_b$, respectively.
For example, the tuples in $g_3$ 
satisfy $a \leq 4$ and $b \leq 6$.
Given a predicate $p_b: b \leq 6$,
\sys determines that it overlaps only 
with $g_3$ and $g_4$, synthesizes $p_a: a \leq 8$, 
and uses $p_a$ to construct a \gsp for 
the index, which could be $\{g_1\}$ or $\{g_3, g_4\}$, 
depending on the \utility values.
During query execution, \sys uses the original predicate $p_b: b \leq 6$ to filter the tuples.

\stitle{Application to arbitrary filters} 
This optimization can be generalized to filters on multiple attributes. 
Given a predicate $p = C_1 \lor C_2 \lor \cdots \lor C_m$, 
\sys examines each conjunctive clause $C_i$ to determine whether it contains an atomic predicate, say $p_b$, 
on an attribute $b$ for which no ANNS index exists. 
If such $p_b$ exists, \sys selects an index 
whose attribute, say $a$, is most correlated with $b$ 
but is not involved in $C_i$ to process $p_b$. 
Then, we synthesize a predicate $p_a$ from $p_b$ and 
replace $p_b$ with $p_a$ in $C_i$, 
resulting in a new clause $C_i'$. 
Finally, we construct the updated predicate 
$p' = C_1' \lor C_2' \lor \cdots \lor C_m'$ 
to generate the \gsp using Algorithms~\ref{alg:conjunction}--\ref{alg:disjunction}. 
During query execution, the executor applies the original predicate $p$.

%% file: implementation.tex
\section{Implementation}
\label{sec:impl}

We implement a prototype of \sys in C++. 
It allows users to load a collection of data objects  
along with their attributes and embeddings as a database relation. 
Each tuple in the relation is automatically assigned a unique 
integer primary key, 
and a primary index is created to locate tuples by this key. 
The primary index is implemented as either a \btree (for update support) 
or an array (for read-only workloads), with \btree as the default. 

Users can selectively build either tree-based 
or hash-based indexes on a subset of attributes to efficiently support filtered ANNS. 
The tree-based index is implemented as a multi-way tree on an attribute~\cite{iRangeGraph:10.1145/3698814,DIGRA}, 
where each node represents a subrange of the attribute’s values. 
For each node, \sys builds a Vamana graph~\cite{DiskANN:conf/nips/SubramanyaDSKK19} 
over the tuples contained in that node’s subrange. 
The graph is constructed on the primary keys of these tuples, 
with primary keys serving as vertex IDs. 
During graph search, \sys uses the primary keys to access 
the corresponding tuple’s attributes and vectors using the primary index. 
The fan-out factor of a tree-based index is configurable and set 
to 2 by default. 
For hash-based indexes, \sys also uses Vamana graphs 
as the \pgraph{s}. 
The updates to tree-based indexes can be handled by 
an existing method~\cite{DIGRA} and updates 
to hash-based indexes can be handled by the 
existing methods for updating \pgraph{s}~\cite{FreshDiskANN:journals/corr/abs-2105-09613}.

Users can issue similarity queries with filters. 
For each query, \sys generates a \gsp, 
searches each \pgraph in this plan to obtain 
intermediate top-$K$ results, and merges them to get the final top-$K$ results. 
\sys employs a best-first search strategy optimized for filtered ANNS, 
referred to as \emph{out-of-range search}~\cite{iRangeGraph:10.1145/3698814,RWalks-SIGMOD25}. 
When exploring a vertex’s neighbors, this strategy considers 
all neighbors (including those that do not satisfy the filter) 
as candidates for further expansion, while using a separate 
queue for maintaining the top-$K$ results passing the filter. 

%% file: experiments.tex
\section{Evaluation}
\label{sec:eval} 
We evaluate \sys to answer the following research questions:
\begin{itemize}
    \item What are the end-to-end performance benefits of \sys for filtered ANNS workloads with conjunctive predicates? (Section~\ref{sec:exp-conjunction})
    \item What are the end-to-end performance benefits of \sys for filtered ANNS workloads with mixed conjunctive and disjunctive predicates? (Section~\ref{sec:exp-disjunction})
    \item What are the performance benefits of the index-borrowing optimization under different levels of attribute correlation? (Section~\ref{sec:exp-index-borrowing})
    \item How does the $\alpha$ parameter in \utility impact the performance of \sys? (Section~\ref{sec:exp-parameter})
    \item How do different index sizes affect query performance, index construction time, and memory overhead? (Section~\ref{sec:exp-index-overhead})
\end{itemize}

\subsection{Experimental setup} 
We run all experiments in a machine that includes 
an AMD EPYC 9454 CPU and 256 GB local DRAM. 
We use Ubuntu 20.04 as the OS and 
16 threads for all baselines.

\begin{figure*}[!t]
    \centering

    \includegraphics[width=0.8\linewidth]{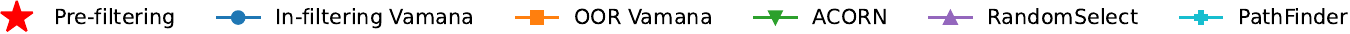}
    \vspace{2px}


    \begin{minipage}{\textwidth}
        \begin{minipage}{0.24\textwidth}
          \includegraphics[width=\linewidth]{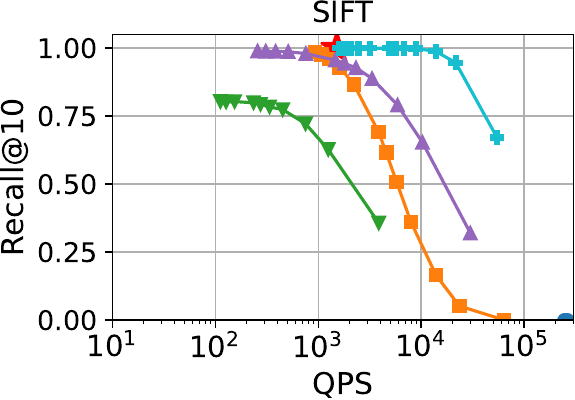}
        \end{minipage}
        \begin{minipage}{0.24\textwidth}
          \includegraphics[width=\linewidth]{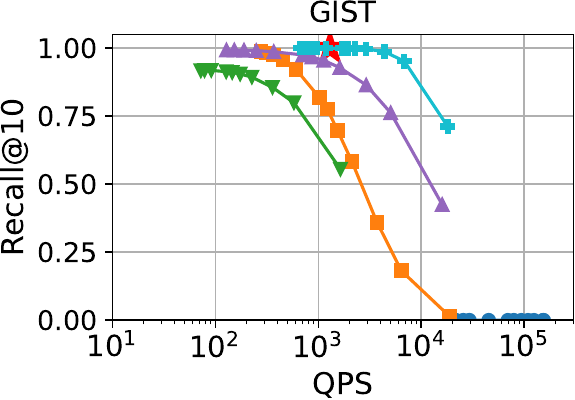}
        \end{minipage}
        \begin{minipage}{0.24\textwidth}
          \includegraphics[width=\linewidth]{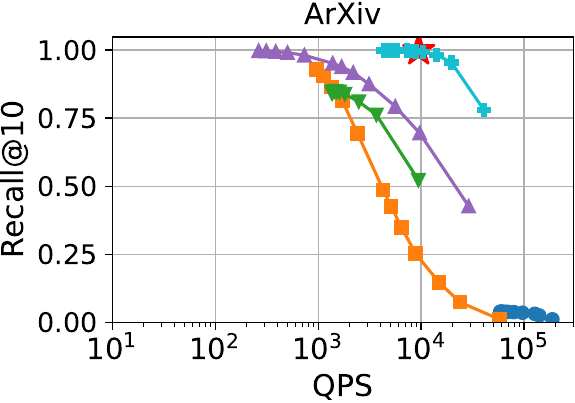}
        \end{minipage}
        \begin{minipage}{0.24\textwidth}
          \includegraphics[width=\linewidth]{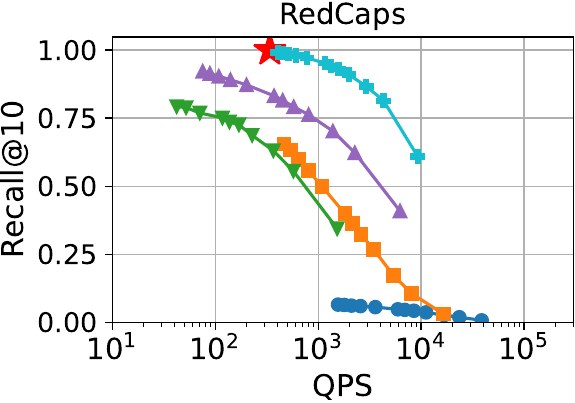}
        \end{minipage}
        \vspace{-2mm}
        \subcaption{Low-selectivity queries (0.1\%--1\%)}
        \label{fig:exp-conjunction-low}
        \vspace{2mm}
    \end{minipage}

    \begin{minipage}{\textwidth}
        \begin{minipage}{0.24\textwidth}
          \includegraphics[width=\linewidth]{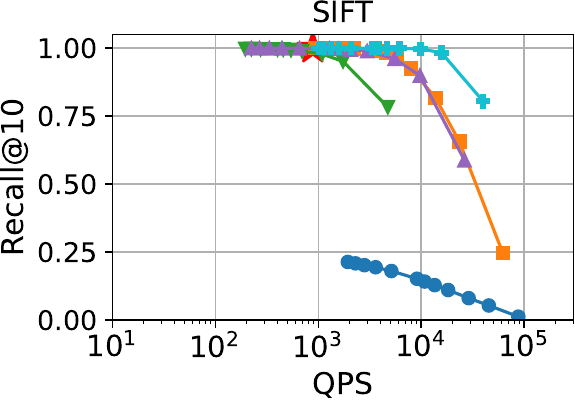}
        \end{minipage}
        \begin{minipage}{0.24\textwidth}
          \includegraphics[width=\linewidth]{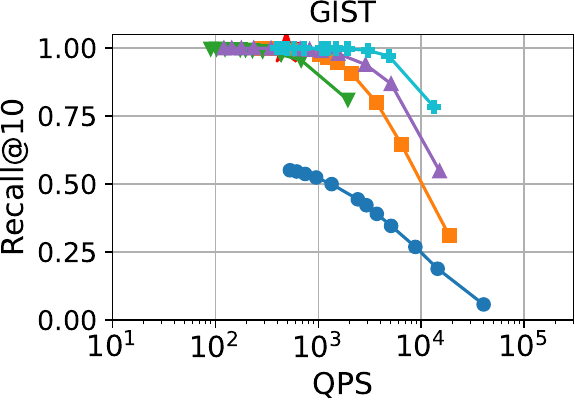}
        \end{minipage}
        \begin{minipage}{0.24\textwidth}
          \includegraphics[width=\linewidth]{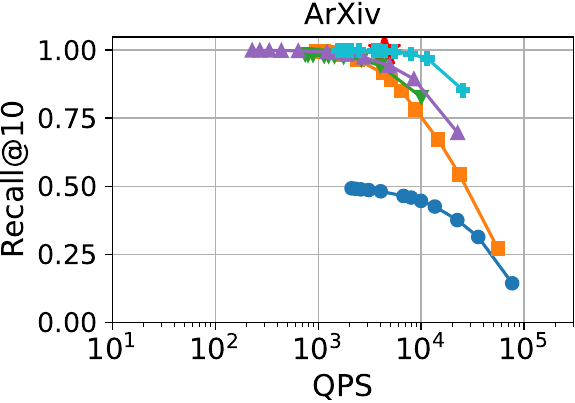}
        \end{minipage}
        \begin{minipage}{0.24\textwidth}
          \includegraphics[width=\linewidth]{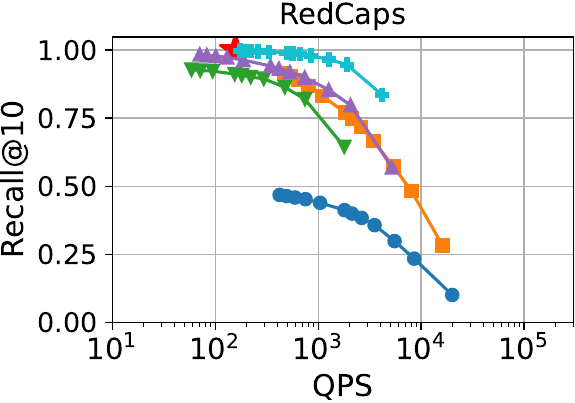}
        \end{minipage}
        \vspace{-2mm}
        \subcaption{Medium-selectivity queries (1\%--10\%) }
        \label{fig:exp-conjunction-medium}
        \vspace{2mm}
    \end{minipage}

    \begin{minipage}{\textwidth}
        \begin{minipage}{0.24\textwidth} 
          \includegraphics[width=\linewidth]{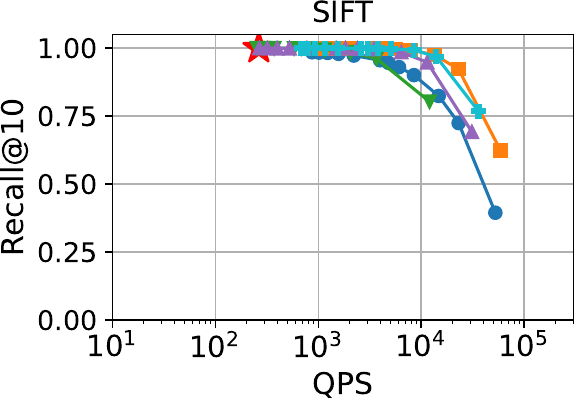}
        \end{minipage}
        \begin{minipage}{0.24\textwidth} 
          \includegraphics[width=\linewidth]{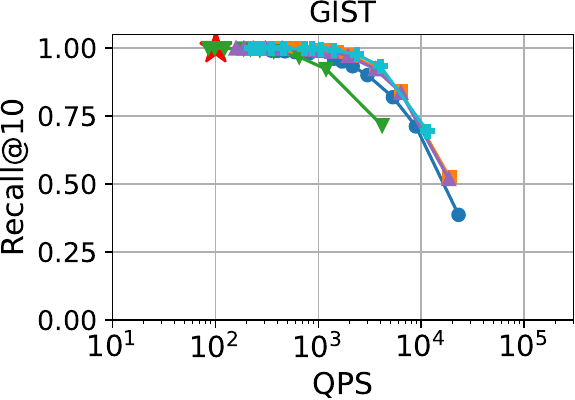}
        \end{minipage}
        \begin{minipage}{0.24\textwidth} 
          \includegraphics[width=\linewidth]{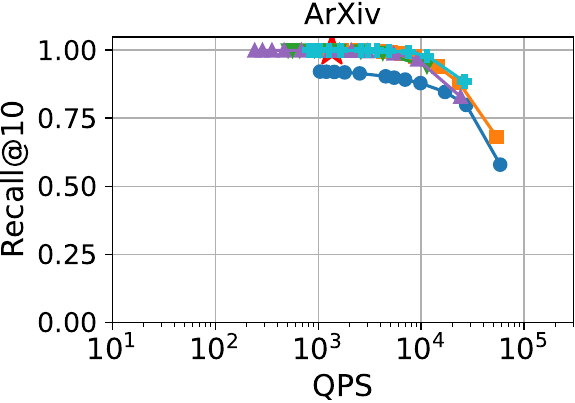}
        \end{minipage}
        \begin{minipage}{0.24\textwidth} 
          \includegraphics[width=\linewidth]{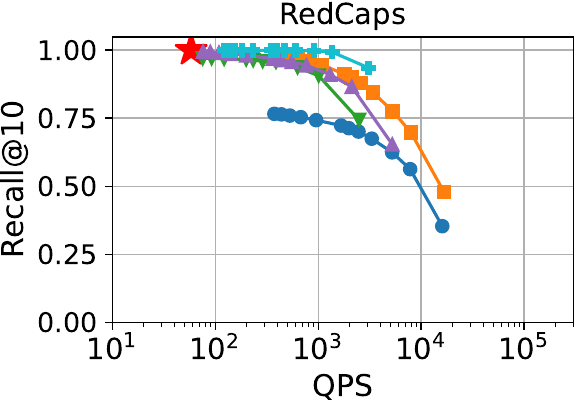}
        \end{minipage}
        \vspace{-2mm}
        \subcaption{High-selectivity queries (10\%--100\%)}
        \label{fig:exp-conjunction-high}
        \vspace{2mm}
    \end{minipage}

    \vspace{-5mm}
    \caption{QPS (queries per second) vs. recall on filters with conjunctive predicates}
    \label{fig:exp-conjunction}
    \vspace{-2mm}
\end{figure*}

\begin{figure*}[!t]
    \centering

    \includegraphics[width=0.8\linewidth]{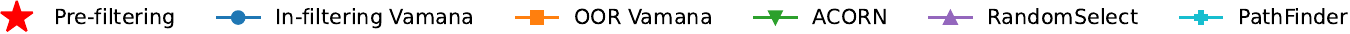}
    \vspace{2px}


    \begin{minipage}{0.33\textwidth}
      \includegraphics[width=0.9\linewidth]{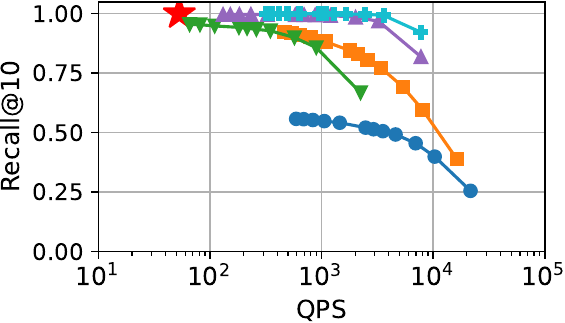}
      \vspace{-2mm}
      \subcaption{Predicates with one attribute}
      \label{fig:exp-conjunction-one}
    \end{minipage}
    \begin{minipage}{0.33\textwidth}
      \includegraphics[width=0.9\linewidth]{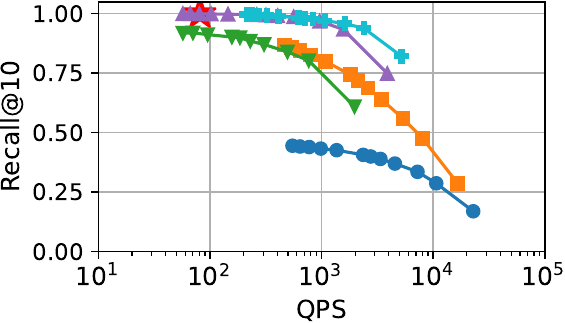}
      \vspace{-2mm}
      \subcaption{Predicates with two attributes}
      \label{fig:exp-conjunction-two}
    \end{minipage}
    \begin{minipage}{0.33\textwidth}
      \includegraphics[width=0.9\linewidth]{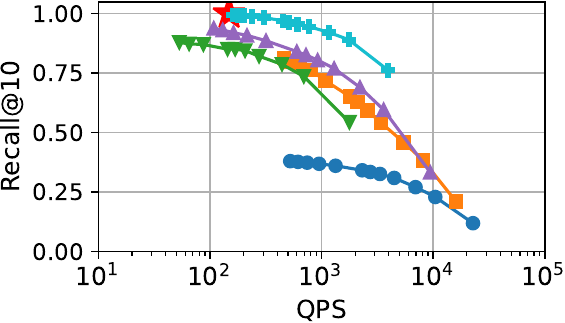}
      \vspace{-2mm}
      \subcaption{Predicates with three attributes}
      \label{fig:exp-conjunction-three}
    \end{minipage}

    \vspace{-3mm}
    \caption{Evaluation on filters with conjunctive predicates that involve different number of attributes (\redcaps)}
    \label{fig:exp-conjunction-count}
    \vspace{-3mm}
\end{figure*}

\begin{figure*}[!t]
    \centering

    \includegraphics[width=0.8\linewidth]{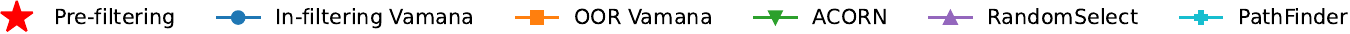}
    \vspace{2px}


    \begin{minipage}{\textwidth}
        \begin{minipage}{0.24\textwidth}
          \includegraphics[width=\linewidth]{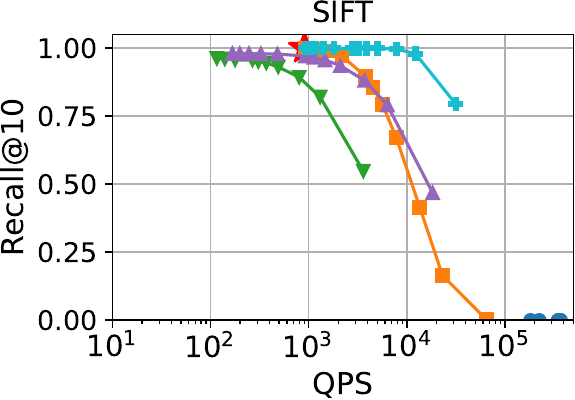}
        \end{minipage}
        \begin{minipage}{0.24\textwidth}
          \includegraphics[width=\linewidth]{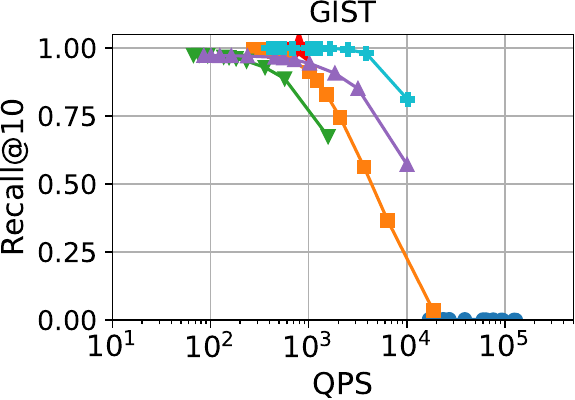}
        \end{minipage}
        \begin{minipage}{0.24\textwidth}
          \includegraphics[width=\linewidth]{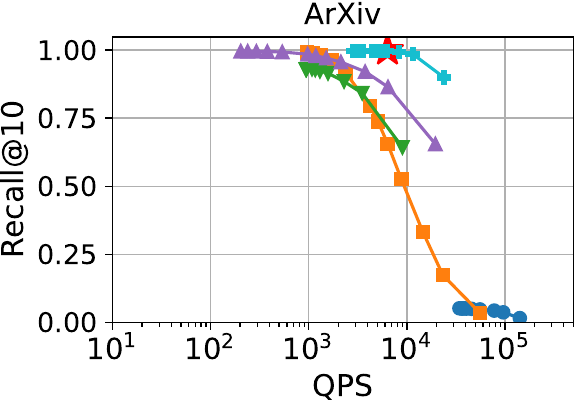}
        \end{minipage}
        \begin{minipage}{0.24\textwidth}
          \includegraphics[width=\linewidth]{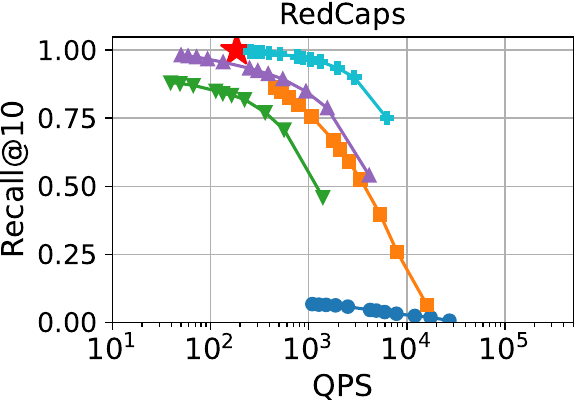}
        \end{minipage}
        \vspace{-2mm}
        \subcaption{Low-selectivity queries (0.1\%--1\%)}
        \label{fig:exp-disjunction-low}
        \vspace{2mm}
    \end{minipage}

    \begin{minipage}{\textwidth}
        \begin{minipage}{0.24\textwidth}
          \includegraphics[width=\linewidth]{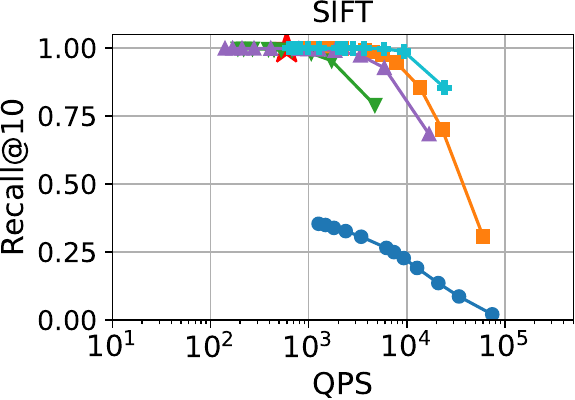}
        \end{minipage}
        \begin{minipage}{0.24\textwidth}
          \includegraphics[width=\linewidth]{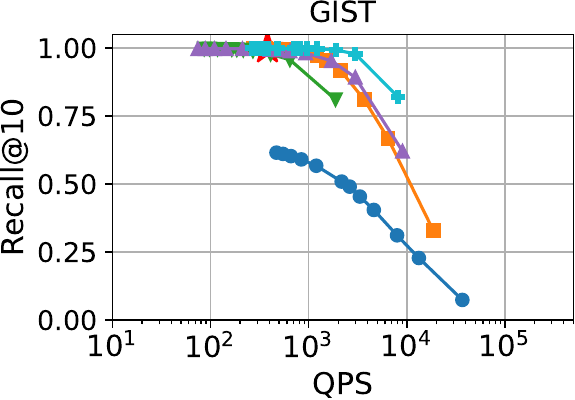}
        \end{minipage}
        \begin{minipage}{0.24\textwidth}
          \includegraphics[width=\linewidth]{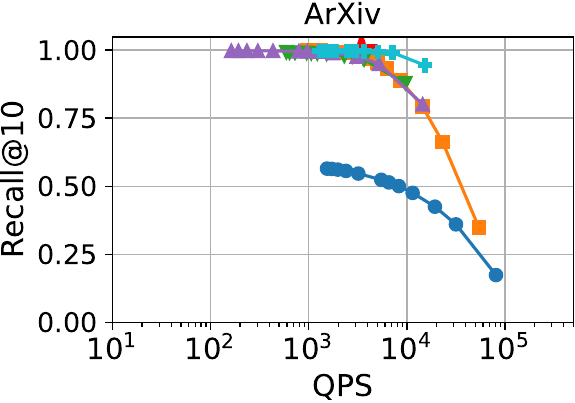}
        \end{minipage}
        \begin{minipage}{0.24\textwidth}
          \includegraphics[width=\linewidth]{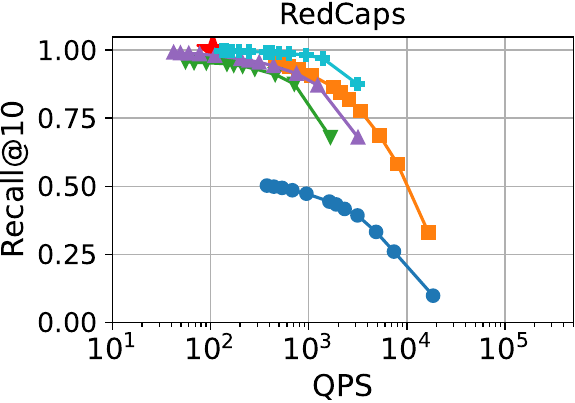}
        \end{minipage}
        \vspace{-2mm}
        \subcaption{Medium-selectivity queries (1\%--10\%)}
        \vspace{2mm}
    \end{minipage}

    \begin{minipage}{\textwidth}
        \begin{minipage}{0.24\textwidth} 
          \includegraphics[width=\linewidth]{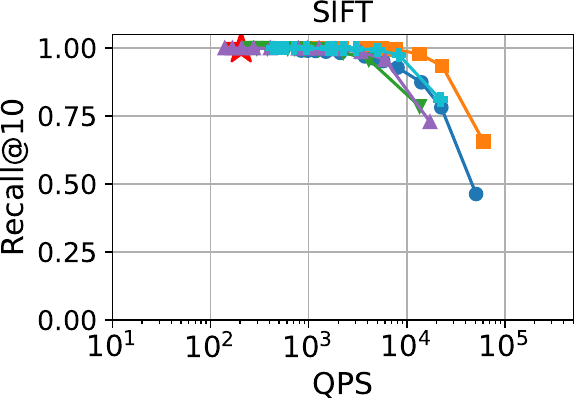}
        \end{minipage}
        \begin{minipage}{0.24\textwidth} 
          \includegraphics[width=\linewidth]{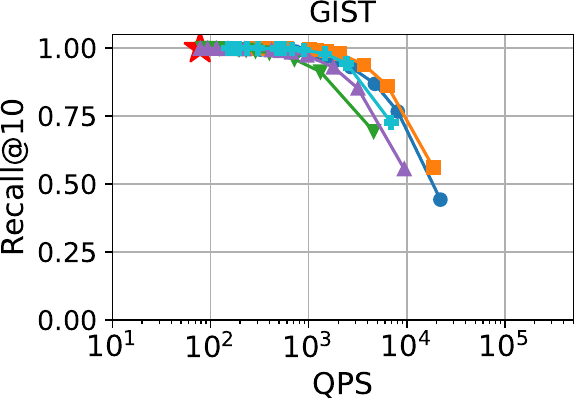}
        \end{minipage}
        \begin{minipage}{0.24\textwidth} 
          \includegraphics[width=\linewidth]{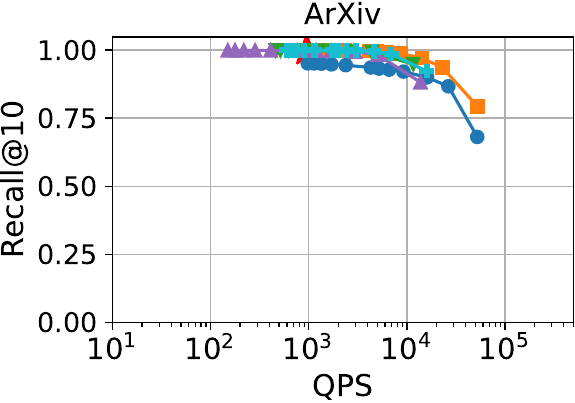}
        \end{minipage}
        \begin{minipage}{0.24\textwidth} 
          \includegraphics[width=\linewidth]{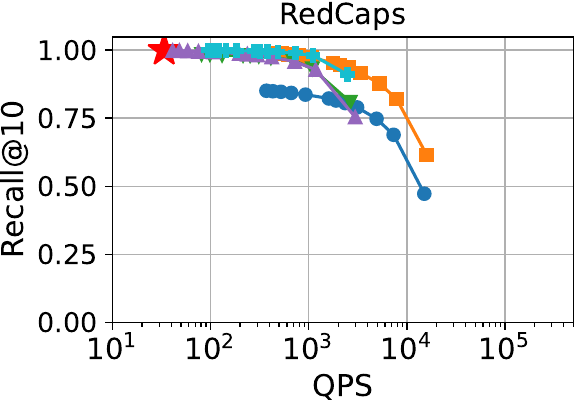}
        \end{minipage}
        \vspace{-2mm}
        \subcaption{High-selectivity queries (10\%--100\%)}
        \vspace{2mm}
    \end{minipage}

    \vspace{-5mm}
    \caption{QPS vs. recall on filters with mixed disjunctive and conjunctive predicates} 
    \label{fig:exp-disjunction}
    \vspace{-5mm}
\end{figure*}

\stitle{Baselines} 
We choose existing approaches that support filters comprising conjunctions and disjunctions on numerical and categorical attributes, as well as a baseline we implement that naïvely leverages attribute-specific indexes.
Specifically, we include two basic search strategies:
1) pre-filtering search, which directly computes distances to all data objects that pass the filter without using any ANNS index (denoted as \emph{\prefilter}), and
2) in-filtering search, which performs best-first search on a Vamana graph~\cite{DiskANN:conf/nips/SubramanyaDSKK19} but only explores vertices that satisfy the filter (denoted as \emph{\infilter}). We omit the post-filtering strategy because its performance is dominated by the out-of-range search baseline~\cite{RWalks-SIGMOD25, iRangeGraph:10.1145/3698814}, which explores all neighbors in a Vamana graph while maintaining a separate queue for the top-$K$ candidates that satisfy the filter, denoted as \emph{\oor}.
We choose the Vamana graph for all baselines since it is widely adopted in industry~\cite{DiskANN:conf/nips/SubramanyaDSKK19, milvus, distributedann} and is also used by \sys. 
We use \acorn~\cite{ACORN:journals/pacmmod/PatelKGZ24} as another baseline, which selectively materializes additional edges to preserve graph density under selective filters and adopts the in-filtering search strategy.

We additionally implement a new baseline, \emph{\randomselect}, which converts the filter predicate into DNF, gets the top-$K$ results for each conjunctive clause separately, and combines the results.
For each conjunctive clause, it randomly selects one of the attribute-specific indexes referenced by the conjunction and applies an existing search algorithm on that index. 
For the tree-based index, we adopt iRangeGraph~\cite{iRangeGraph:10.1145/3698814} as it supports a conjunctive predicate that involves the attribute this index is built on.  
For the hash-based index, we search all \pgraph{s} involved in the predicate using the out-of-range search and merge their results.
All baselines are implemented in the \sys codebase for a fair comparison.

\stitle{Configurations} 
For the Vamana graph, we set the parameter for robust pruning to $\alpha = 1.2$~\cite{DiskANN:conf/nips/SubramanyaDSKK19}.
For the tree-based index, we use a fan-out factor of 2 and a tree height of 7. We build attribute-specific indexes for all attributes of the datasets. 
In Section~\ref{sec:exp-index-overhead}, we vary the tree height and the number of available indexes to evaluate their impact.
\sys and \randomselect use the same attribute-specific indexes. 
We use $\alpha=0.4$ for estimating the \utility (Equation~\ref{eq:utility}) 
For ACORN, we use $M_{\beta} = 128$ and $\gamma=80$. 
We report the tradeoff between the query throughput and the average recall@10 by varying the search queue length of the best-first search from 10 to 3,000 for all approaches.  

\begin{table}[!t]
\caption{Datasets used in the evaluation}
\label{tbl:datasets}
\vspace{-3mm}
\small
\begin{tabular}{|c|cccc|}
\hline
        & \#Rows & \#Dim. & Numeric Attr.                                                             & Categorical Attr.                                                                  \\ \hline
ArXiv   & 132K   & 768   & \begin{tabular}[c]{@{}c@{}}publication year\\ citation count\end{tabular} & \begin{tabular}[c]{@{}c@{}}classification\\ (142 distinct values)\end{tabular}            \\ \hline
RedCaps & 6.9M   & 512   & \begin{tabular}[c]{@{}c@{}}post time\\ upvote count\end{tabular}          & \begin{tabular}[c]{@{}c@{}}subreddit category\\ (350 distinct values)\end{tabular} \\ \hline
SIFT    & 1M     & 128    & 4 decimals                                                              & N/A                                                                                \\ \hline
GIST    & 1M     & 960    & 4 decimals                                                               & N/A                                                                                \\ \hline
\end{tabular}
\vspace{-6mm}
\end{table}

\stitle{Benchmarks} 
We evaluate \sys and all baselines on four real-world datasets, summarized in Table~\ref{tbl:datasets}. 
\arxiv is a dataset of research papers from the arXiv repository, where vector embeddings are generated from paper abstracts~\cite{arxiv}.
\redcaps is an image dataset with attributes extracted from Reddit~\cite{redcaps}.
We use two numeric and one categorical attribute for both datasets. 
\sift and \gist are standard benchmarks for evaluating ANNS algorithms~\cite{sift}.
For each of them, we create four attributes of decimal numbers, with the first two correlated and the last two independent. 
To generate correlated attributes, 
we sample one attribute $a$ from a standard Gaussian distribution and then generate another attribute $b$ using the formula $b = a + k \times norm$, where $norm$ follows the standard Gaussian distribution. 
The parameter $k$ controls the correlation level.
We set $k = 0.5$ for \sift and $k = 0.1$ for \gist to introduce different levels of correlation. 
The values of the last two attributes are uniformly sampled  
from the value range [0, 1000]. 
To build the Vamana graph and \acorn, we set the max neighbor degree $M$ to 32 for SIFT and 64 for all other datasets. 

To generate similarity queries, we use the queries provided with the datasets (for \sift and \gist) or extract a random sample of data objects from the dataset and exclude them from the dataset (for \arxiv and \redcaps).
Filters are generated by combining two templates of atomic predicates on different attributes using conjunctions and disjunctions.
The first template is $min \leq a \leq max$, where $a$ is a numeric attribute and $min$/$max$ values are configurable.
The second one is $a \text{ IN } C$, where $a$ is a categorical attribute and $C$ is a set of categorical values.
For each test, we generate 1K filtered similarity queries and scan the dataset to obtain the ground truth. 

\subsection{Performance on Conjunctive Predicates}
\label{sec:exp-conjunction} 
In this subsection, we evaluate \sys and all baselines under filtered ANNS workloads with conjunctive predicates. 
We generate a conjunctive predicate by randomly creating atomic predicates over all attributes and combining two or three of them using conjunctions. We group these predicates by selectivity and the number of attributes involved and evaluate them separately.

\stitle{Varying selectivity level} 
For each dataset, we construct three groups of filters with different selectivity levels: 
low (0.1\%--1\%), medium (1\%--10\%), and high (10\%--100\%), with each group having 1K queries. 
Figure~\ref{fig:exp-conjunction} presents the QPS (queries per second) and recall curves across all selectivity levels 
and datasets. 
We observe that \sys can achieve nearly 1.0 
recall across all workloads, 
while all baselines except \prefilter fall short.  
\prefilter always has perfect recall because it directly computes distances without using ANNS indexes; however, \sys 
has substantially higher QPS than \prefilter. 
\sys consistently outperforms all other baselines under low- and medium-selectivity workloads, 
and achieves comparable or better performance under high selectivity. 
This is because \sys can effectively utilize 
attribute-specific indexes by quickly constructing  
an efficient \gsp using the 
\utility metric and the optimization algorithms. 
For example, compared to \randomselect, the strongest baseline for low- and medium-selectivity workloads, \sys achieves 18.4$\times$ higher QPS at recall=0.9 (i.e., for RedCaps in Figure~\ref{fig:exp-conjunction-low}). 
\randomselect performs better than other baselines because it can leverage attribute-specific indexes, 
whereas \infilter suffers from low recall due to the sparsity of filtered graphs under low- and medium-selectivity workloads. 
The optimization overhead of \sys (included in the reported QPS and recall curves) is small, accounting for only 0.12\%–5.43\% of the total end-to-end execution time across the four datasets.
\redcaps exhibits the lowest relative overhead (0.12\%) due to its large number of data objects, whereas \arxiv shows the highest overhead (5.43\%).

\stitle{Varying the number of attributes}
Next, we categorize the conjunctive predicates 
in the \redcaps dataset
by the number of attributes involved (two or three)
and additionally construct a group of single-attribute predicates.
Each group contains 1K queries, with one-third of the queries corresponding to each selectivity level. 
Figure~\ref{fig:exp-conjunction-count} shows that 
\sys achieves a better tradeoff between QPS and recall 
than all baselines. 
\sys outperforms \randomselect on single-attribute predicates because its cost-based optimization framework can generate an execution plan 
that searches the tree-based or hash-based index more efficiently than the baseline approaches used by \randomselect. 
These results show that the cost-based optimization in \sys not only benefits multi-attribute filters but also improves performance for single-attribute filters.

\subsection{Performance on Mixed Conjunctive and Disjunctive Predicates}
\label{sec:exp-disjunction} 
Now we evaluate filtered ANNS workloads with mixed conjunctive 
and disjunctive predicates. 
We generate such a predicate by first generating two conjunctive or atomic predicates (using the method from Section~\ref{sec:exp-conjunction}) and connecting them 
using a disjunction. 
For each dataset, we evaluate three groups of predicates, each corresponding to a selectivity level. 
Figure~\ref{fig:exp-disjunction} shows 
that \sys can achieve almost 1.0 recall for all workloads while the baselines (except \prefilter) cannot 
and \sys has a stronger QPS and recall tradeoff than all baselines for the low- and medium-selectivity workloads. 
Specifically, \sys has 9.8$\times$ higher QPS than 
\randomselect, the strongest baseline, at recall 0.95 (i.e., for RedCaps in Figure~\ref{fig:exp-disjunction-low}). 
The optimization overhead of \sys is also small in these tests, 
accounting for 0.11\%–3.33\% of the total execution time across the four datasets.
\sys shows slightly lower performance than \oor for high-selectivity workloads when the recall is below 0.9, because \sys may search \pgraph{s} across different indexes, whereas \oor searches only a single graph.

%
%

\begin{figure*}[!t]
  \centering
  \begin{tabular}{cc}
      \begin{minipage}{0.74\textwidth}
          \centering

          \begin{minipage}{4.3cm}
            \includegraphics[width=\linewidth]{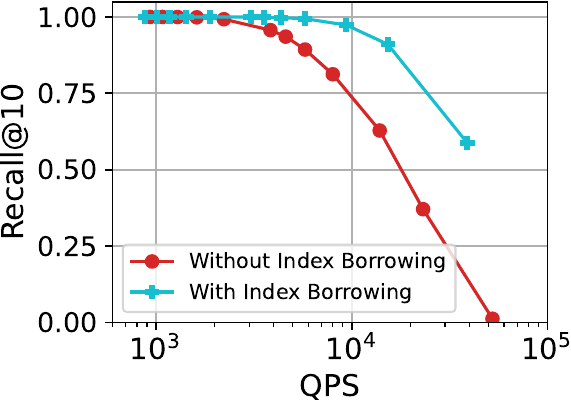}
            \subcaption{High correlation (k=0.05)}
          \end{minipage}
          \begin{minipage}{4.3cm}
            \includegraphics[width=\linewidth]{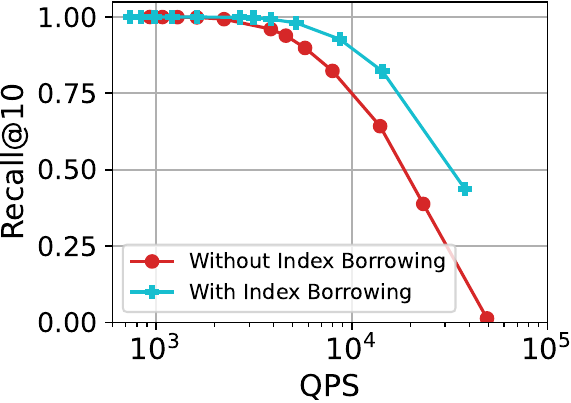}
            \subcaption{Medium correlation (k=0.1)}
          \end{minipage}
          \begin{minipage}{4.3cm}
            \includegraphics[width=\linewidth]{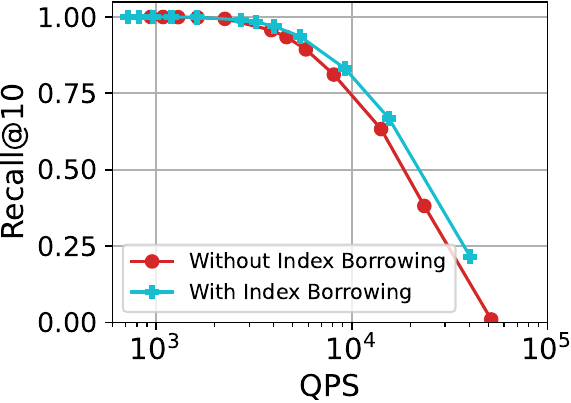}
            \subcaption{Low correlation (k=0.2)}
          \end{minipage}

          \vspace{-3mm}
          \caption{Evaluation of index borrowing under varying attribute correlations (\sift)}
          \label{fig:exp-index-borrowing}
      \end{minipage}
      
      \begin{minipage}{0.25\textwidth}
          \centering
          \begin{minipage}{4.3cm}
            \includegraphics[width=\linewidth]{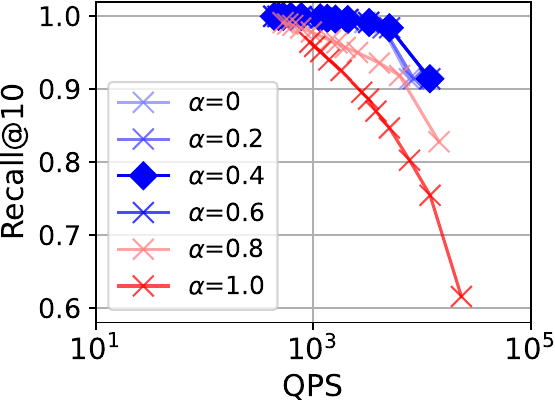}
          \end{minipage}
  
          \vspace{-3px}
          \caption{The performance impact of the $\alpha$ parameter}
          \label{fig:exp-alpha-ablation}
      \end{minipage}
  \end{tabular}
\end{figure*}

\begin{figure*}
    \centering
    \includegraphics[width=0.8\linewidth]{experiment-figures/fig1-conjunctive/fig1-legend.pdf}
    \vspace{2px}

    \def\mylen{0.195\textwidth}

    \begin{minipage}{\mylen}
      \includegraphics[width=\linewidth]{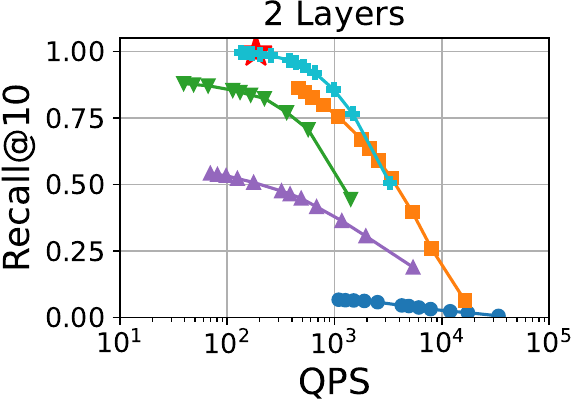}
    \end{minipage}
    \begin{minipage}{\mylen}
      \includegraphics[width=\linewidth]{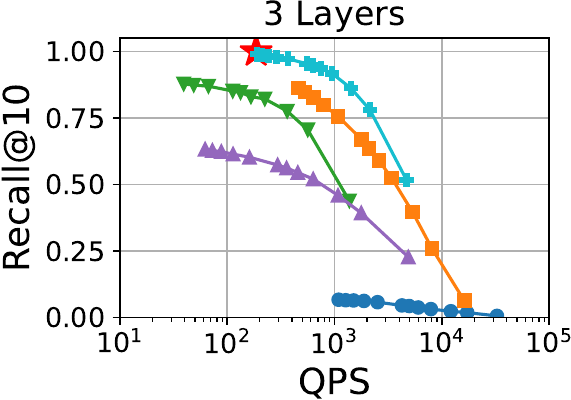}
    \end{minipage}
    \begin{minipage}{\mylen}
      \includegraphics[width=\linewidth]{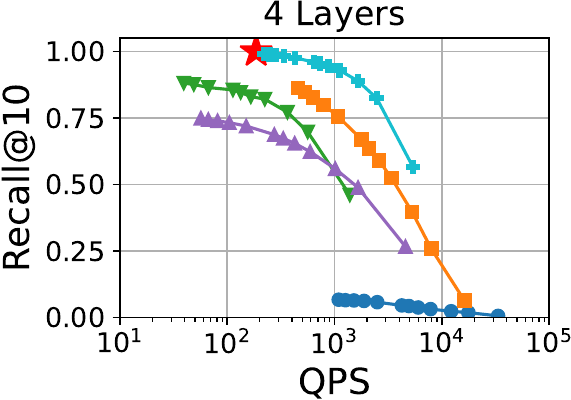}
    \end{minipage}
    \begin{minipage}{\mylen}
      \includegraphics[width=\linewidth]{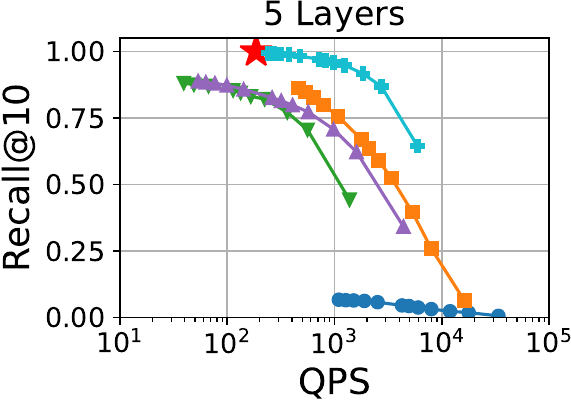}
    \end{minipage}
    \begin{minipage}{\mylen}
      \includegraphics[width=\linewidth]{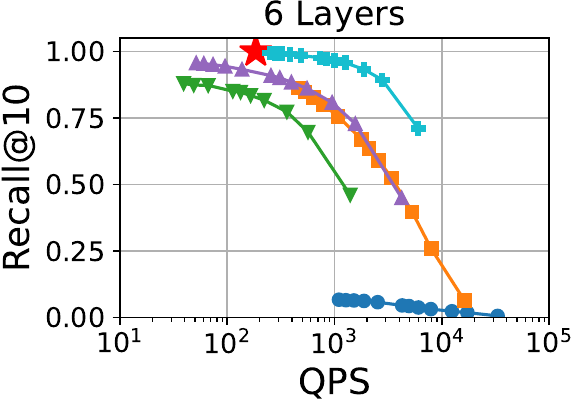}
    \end{minipage}

    \caption{Performance impact of using different numbers of layers for the tree-based indexes (\redcaps)}
    \label{fig:exp-index-layer}
    \vspace{-0mm}
\end{figure*}

\begin{figure}
    \centering
    \includegraphics[width=0.85\linewidth]{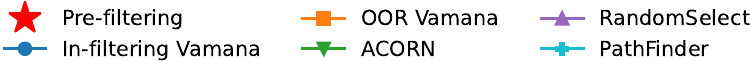}
    \vspace{4px}
    \def\mylen{0.48\linewidth}

    \begin{minipage}{\mylen}
      \includegraphics[width=\linewidth]{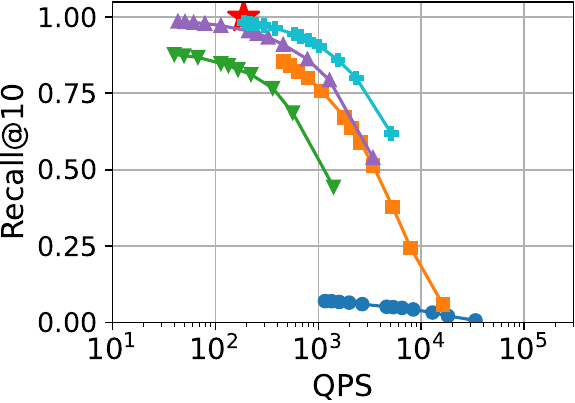}
      \subcaption{Two indexes available}
    \end{minipage}
    \begin{minipage}{\mylen}
      \includegraphics[width=\linewidth]{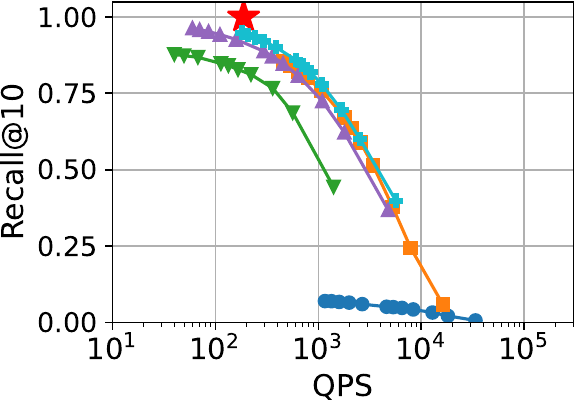}
      \subcaption{One index available}
    \end{minipage}

    \caption{Varying the number of available indexes (\redcaps)}
    \label{fig:exp-index-count}
    \vspace{-5mm}
\end{figure}

\subsection{Performance of Index Borrowing}
\label{sec:exp-index-borrowing}
We next evaluate the performance benefits of the index borrowing optimization under different levels of attribute correlation. 
Recall that we synthesize a pair of correlated attributes for the \sift and \gist datasets using the formula 
$b = a + k \times norm$, where $a$ is an attribute generated from a standard Gaussian distribution, 
and $norm$ is another standard Gaussian variable. 
By varying the value of $k$, we control the strength of the correlation between $a$ and $b$. 
In this experiment, we build an attribute-specific index on one of the correlated attributes and issue queries with filters on the other. 
We generate 1K queries, with one-third of the queries corresponding to each selectivity level. 

Figure~\ref{fig:exp-index-borrowing} shows the QPS and recall curves 
for \sys with and without the index borrowing optimization under 
three different $k$ values for the \sift dataset. 
In the setting without index borrowing, \sys searches the root graph 
using the out-of-range search strategy.  
We observe that the index borrowing optimization significantly 
improves performance when the attributes are highly correlated 
(e.g., $k=0.05$), and that the performance gain gradually diminishes 
as the correlation weakens. 
Specifically, the index borrowing optimization improves QPS 
by up to 2.44$\times$ at recall=0.95. 
The results for the \gist dataset exhibit the same trend 
and are omitted due to space limitations.

\subsection{Performance Impact of the $\alpha$ Parameter}
\label{sec:exp-parameter}
We evaluate the performance impact of the $\alpha$ parameter 
in the \utility function. 
As defined in Equation~\ref{eq:utility}, $\alpha$ controls 
the penalty for searching multiple \pgraph{s}: 
a larger $\alpha$ value discourages the optimizer from 
selecting multiple graphs. 
We study its performance impact 
by evaluating predicates on the categorical attribute 
of the \redcaps dataset with varying $\alpha$ value. 
This attribute includes 350 distinct values. 
We generate 1K predicates, each randomly 
selecting up to 30 attribute values 
(i.e., using the $a \text{ IN } C$ template). 
In this setting, \sys needs to choose between using the root graph 
or the hash-based index that searches multiple \pgraph{s}. 
Figure~\ref{fig:exp-alpha-ablation} shows that 
$\alpha=1.0$ results in the lowest performance, 
as it over-penalizes the use of multiple graphs and forces 
\sys to use the single root graph. 
Conversely, $\alpha=0.0$ does not achieve the best performance 
either, as it causes the optimizer to always use 
the hash-based index and search too many graphs. 
We find that $\alpha=0.4$ and $\alpha=0.6$ offer the best tradeoff 
between recall and QPS; therefore, we set 
$\alpha=0.4$ as the default value in our experiments.

\subsection{Performance Impact of Varying Index Sizes}
\label{sec:exp-index-overhead} 
We now evaluate the performance impact of different 
index sizes by varying 
1) the number of layers in the tree-based indexes 
and 2) the number of indexes available. 

Since our default configuration builds 7 layers of tree-based indexes, we vary this number from 6 to 2 and report the query performance and indexing overhead.
Figure~\ref{fig:exp-index-layer} shows the QPS and recall curves for the mixed conjunctive and disjunctive predicates at the low-selectivity level for the \redcaps dataset.
For the same QPS, the recall of both \sys and \randomselect drops as we reduce the number of layers for the tree-based indexes.
The peak recall of \randomselect drops significantly, from 0.98 to 0.54, while \sys maintains a recall close to 1.0.
Even with 3 layers, \sys still achieves a better QPS–recall tradeoff than all other baselines. For example, at a recall of 0.85, \sys has 3.1$\times$ higher QPS than \oor.
When using only 2 layers, \sys shows a similar performance to \oor for recall below 0.7, but has a stronger tradeoff between QPS and recall 
and the maximum recall in other cases. 

Table~\ref{tbl:index-time} and Table~\ref{tbl:index-size} report the index construction time and index sizes for all datasets when building indexes for all attributes, respectively.
Although \sys takes a longer index construction time than the Vamana graph, its construction time is substantially smaller than that of \acorn.
\sys consumes more memory than both \acorn and Vamana, 
but reducing the number of layers from 7 to 2 cuts memory usage by 3.7$\times$ on average while still allowing \sys to outperform all baselines on query performance. 
An interesting direction for future work is to compress attribute-specific indexes to reduce memory overhead.

Next, we vary the number of indexes available on the \redcaps dataset.
Since \redcaps has three attributes, we construct six index configurations,
each having indexes available on two or one attribute. 
We evaluate the mixed disjunctive and conjunctive predicates
from the low-selectivity group under each configuration
and aggregate the QPS and recall results by the number of indexes available.
Figure~\ref{fig:exp-index-count} show that while the performance gains of \sys decrease
as fewer indexes are available, it consistently outperforms all baselines. 
For example, when two indexes are available, \sys has 1.82$\times$ higher 
QPS than \randomselect at recall=0.95. 

\begin{table}[!t]\centering
\caption{Time to Index (s)}\label{tab:indexing-time}
\label{tbl:index-time}
\vspace{-3mm}
\begin{tabular}{lrrrrr}\toprule
&SIFT &GIST &ArXiv &RedCaps \\\midrule
ACORN &684.8 &5633.2 &86.3 &31483.3 \\
Vamana &12.9 &162.3 &4.0 &389.1 \\
\textbf{\sys} -- 2 layers &68.1 &762.8 &8.7 &1147.5 \\
\textbf{\sys} -- 3 layers &110.3 &1235.1 &13.8 &1830.5 \\
\textbf{\sys} -- 4 layers &143.8 &1599.6 &17.4 &2407.8 \\
\textbf{\sys} -- 5 layers &169.2 &1865.2 &20.2 &2887.9 \\
\textbf{\sys} -- 6 layers &190.3 &2052.1 &22.1 &3274.4 \\
\textbf{\sys} -- 7 layers &208.1 &2187.4 &23.4 &3548.9 \\
\bottomrule
\end{tabular}
\end{table}

%% file: related.tex
\section{Related Work}
\label{sec:related} 
We discuss the related work on ANNS, filtered ANNS, 
and access path selection in vector databases. 

\stitle{ANNS} 
There has been extensive research on ANNS indexes, which can be broadly categorized into hashing-based~\cite{LSH-SIGMOD09, DB-LSH}, clustering-based~\cite{SPANN::conf/nips/ChenZWLLLYW21, ScANN:icml/GuoSLGSCK20}, and graph-based~\cite{HNSW:journals/pami/MalkovY20, NSW:journals/is/MalkovPLK14, NSG:journals/pvldb/FuXWC19, DiskANN_Out_of_Date:conf/nips/SubramanyaDSKK19} approaches.
We focus on graph-based methods because they provide an excellent tradeoff between QPS and recall~\cite{NSG:journals/pvldb/FuXWC19, HNSW:journals/pami/MalkovY20, DiskANN_Out_of_Date:conf/nips/SubramanyaDSKK19}, and have been adopted in nearly all modern vector databases~\cite{Milvus:wang2021milvus, Analyticdb:wei2020analyticdb, pgvector, chromadb, lancedb}. 
Among them, HNSW~\cite{HNSW:journals/pami/MalkovY20} and Vamana~\cite{DiskANN_Out_of_Date:conf/nips/SubramanyaDSKK19} 
are the most widely adopted ones. 
\sys adopts the Vamana graph as its \pgraph.


\stitle{Filtered ANNS} 
Graph-based indexes support filtered ANNS using either the in-filtering or post-filtering strategy. 
Recent studies have explored new techniques to 
further improve filtered ANNS performance. 
\acorn~\cite{ACORN:journals/pacmmod/PatelKGZ24} supports arbitrary filters and enhances performance by materializing additional edges and utilizing 2-hop search. 
Two recent works~\cite{HQI:10.1145/3589777, SIEVE-VLDB25} support complex filters by materializing filter-specific graph indexes tailored to a known filter workload, an assumption that \sys does not make. 
Another line of research focuses on attribute-specific indexes.
Several studies have proposed graph-based indexes for range filters on numeric data~\cite{SeRF:10.1145/3639324, iRangeGraph:10.1145/3698814, WindowFilter:10.5555/3692070.3692567, DIGRA, WoW, UNIFY-VLDB25}.
Among these, tree-based indexes are the most popular~\cite{WoW, WindowFilter:10.5555/3692070.3692567, iRangeGraph:10.1145/3698814, DIGRA} and have also been extended to support updates~\cite{DIGRA, WoW}. 
Other papers target label data~\cite{UNG:10.1145/3698822, Filtered-DiskANN:10.1145/3543507.3583552, CAPS:gupta2023capspracticalpartitionindex, NHQ:10.5555/3666122.3666814, RWalks-SIGMOD25}.

\sys differs from these approaches in that it serves as a unified indexing framework that leverages attribute-specific indexes to efficiently handle complex filters with conjunctions and disjunctions.
It supports both numeric and categorical attributes and achieves state-of-the-art performance on filtered ANNS workloads over these data types. 
\sys supports label data using the root graph. 
Extending \sys to support attribute-specific indexes for label data is left for future work.

\stitle{Access path selection} 
Relational databases allow administrators to build indexes on selective attributes and employ an access path selection framework to determine how best to utilize these indexes~\cite{APS:SIGMOD79, APS:SIGMOD17}.
Recent studies on vector databases have also proposed new access path selection frameworks~\cite{Analyticdb:wei2020analyticdb, Milvus:wang2021milvus, APS-vectorDB-DaMon24}, but they focus on choosing among the three basic search strategies: pre-filtering, in-filtering, and post-filtering.
In contrast, \sys focuses on leveraging attribute-specific ANNS indexes to improve the performance of filtered ANNS, and is therefore complementary to the aforementioned frameworks.
 
\begin{table}[!t]\centering
\caption{Index Size (GB)}\label{tab:index-size}
\label{tbl:index-size}
\vspace{-3mm}
\begin{tabular}{lrrrrr}\toprule
&SIFT &GIST &ArXiv &RedCaps \\\midrule
ACORN &0.56 &0.59 &0.08 &4.24 \\
Vamana &0.12 &0.16 &0.02 &1.31 \\
\textbf{\sys} -- 2 layers &0.59 &0.81 &0.08 & 5.11 \\
\textbf{\sys} -- 3 layers &1.07 &1.45 &0.12 & 7.67 \\
\textbf{\sys} -- 4 layers &1.53 &2.08 &0.16 &10.21 \\
\textbf{\sys} -- 5 layers &2.00 &2.70 &0.20 &12.72 \\
\textbf{\sys} -- 6 layers &2.46 &3.30 &0.23 &15.15 \\
\textbf{\sys} -- 7 layers &2.91 &3.89 &0.26 &17.29 \\
\textit{(Raw Vectors)} &\textit{0.48} &\textit{3.58} &\textit{0.35} &\textit{13.24} \\
\bottomrule
\end{tabular}
\end{table}